\documentclass[fleqn,usenatbib]{mnras}

\usepackage{mathptmx}
\usepackage{float}

\usepackage[T1]{fontenc}
\usepackage{ae,aecompl}

\usepackage{graphicx}	
\usepackage{amsmath}	
\usepackage{amssymb}	
\usepackage[normalem]{ulem}
\usepackage{soul}
\usepackage{multirow}
\usepackage{mathtools}
\usepackage{lmodern}

\newcommand{\field}{{\hat \phi}}
\newcommand{\de}[1]{\partial_{#1 }}
\newcommand{\sumnn}{\sum_{j\in \text{NN}(i)}}

\newcommand{\rhoc}{\rho_{c}}
\newcommand{\Rc}{R_{c}}
\newcommand{\Mc}{M_{c}}
\newcommand{\rhovir}{\rho_{\rm vir}}
\newcommand{\Rvir}{R_{\rm vir}}
\newcommand{\Mvir}{M_{\rm vir}}
\newcommand{\sigmavir}{\sigma_{\rm vir}}
\newcommand{\sigmac}{\sigma_{c}}

\newcommand{\dimrho}{\mathrm{M}_{\odot} \mathrm{h}^2/\mathrm{Kpc}^3}
\newcommand{\dimR}{\mathrm{Kpc}/\mathrm{h}}
\newcommand{\dimM}{\mathrm{M}_{\odot} / \mathrm{h}}

\newcommand{\PG}{{\small P-GADGET}3~}
\newcommand{\AG}{{\small AX-GADGET}~}

\newcommand{\MUSIC}{{\small Music}~}
\newcommand{\SUBFIND}{{\small SUBFIND}~}
\newcommand{\AC}{{\small axionCAMB}~}
\newcommand{\citenp}[1]{\citeauthor{#1} \citeyear{#1}}

\newcommand{\PARTII}{(Nori~\&~Baldi, in prep.)}

\renewcommand{\vec}[1]{\mathbf{#1}}

\title[Scaling relations of Fuzzy Dark Matter haloes I]{Scaling relations of Fuzzy Dark Matter haloes I:\\ 
individual systems in their cosmological environment}

\author[M. Nori et al.]{
Matteo Nori,$^{1,2,3}$\thanks{E-mail: matteo.nori3@unibo.it}
and Marco Baldi$^{1,2,3}$
\\
$^{1}$Dipartimento di Fisica e Astronomia, Alma Mater Studiorum - University of Bologna, Via Piero Gobetti 93/2, 40129 Bologna BO, Italy\\
$^{2}$INAF - Osservatorio Astronomico di Bologna, Via Piero Gobetti 93/3, 40129 Bologna BO, Italy\\
$^{3}$INFN - Istituto Nazionale di Fisica Nucleare, Sezione di Bologna, Viale Berti Pichat 6/2, 40127 Bologna BO, Italy
}

\date{Accepted XXX. Received YYY; in original form ZZZ}

\pubyear{2020}
\begin{document}
\label{firstpage}
\pagerange{\pageref{firstpage}--\pageref{lastpage}}
\maketitle

\begin{abstract}
Dark matter models involving a very light bosonic particle, generally known as Fuzzy Dark Matter (FDM), have been recently attracting great interest in the cosmology community, as their wave-like phenomenology would simultaneously explain the longstanding mis-detection of a dark matter particle and help easing the small-scale issues related to the standard Cold Dark Matter (CDM) scenario.
With the present work, we initiate a series of papers aiming at investigating the evolution of FDM structures in a cosmological framework performed with our N-body code \AG, detailing for the first time in the literature how the actual scaling relations between solitonic cores and host haloes properties are significantly affected by the dynamical state, morphology and merger history of the individual systems. In particular, in this first paper we confirm the ability of \AG to correctly reproduce the typical FDM solitonic core and we employ it to study the non-linear evolution of eight FDM haloes in their cosmological context through the zoom-in simulation approach. We find that the scaling relations identified in previous works for isolated systems are generally modified for haloes evolving in a realistic cosmological environment, and appear to be valid only as a limit for the most relaxed and spherically symmetric systems.
\end{abstract}




\section{Introduction}
\label{sec:intro}

Whether in the form of a yet undiscovered particle or phenomenologically arising from a more complex mechanism, a \textit{cold} and \textit{dark} matter (CDM) species is an established ingredient of the standard cosmological paradigm. In fact, the presence of a collisionless type of matter with negligible interaction with the electromagnetic field helps explaining the formation and the dynamical properties of cosmic structures over a large range of scales, from the rotation curves of spiral galaxies \citep[][]{Rubin_Ford_Thonnard_1980,Bosma_1981,Persic96} through the inner dynamics of galaxy clusters \citep[][]{Zwicky_1937,Clowe06} up to the cosmological scales probed with weak gravitational lensing generated by the large-scale matter distribution \citep[][]{Mateo98, Heymans_etal_2013, Planck_2018_Gravitional_Lensing, Hildebrandt_etal_2017}.

The identification of a fundamental CDM particle has been critically elusive for a wide variety of direct and indirect detection experiments \citep[see e.g.][]{Fermi17annih, Danninger17, Buonaura18}, challenging the historical consensus that gathered around the hypothesised dark matter particle arising in the context of the so-called Weakly Interactive Massive Particles \citep[WIMPS, ][]{Jungman95} scenario. 

From a cosmological point of view, a well-defined \textit{abundance} of dark matter with respect to the total cosmic energy budget \citep[$\Omega _{\rm{CDM}}=0.264\pm 0.003$, ][]{Planck18} is required in order to be consistent with the cosmic expansion history and with the observed properties of large scale structures --~as explicitly emerging from the comparison between low-redshift surveys and the angular power spectrum of the Cosmic Microwave Background (CMB) temperature anisotropies that seed the early universe density perturbations \citep[observed e.g. from WMAP and Planck][respectively]{wmap7,Planck18}~-- but no specific \textit{mass} range is enforced on the dark matter particle itself.

Therefore, to better substantiate the long-standing dark matter particle mis-detection, the scientific community efforts in the hunt for direct dark matter observations has been shifting from the GeV/$c^2$ mass range of the WIMPs towards lighter candidates. A well-motivated dark matter candidate in such lower mass range is the axion particle arising from the CP-symmetry break in Quantum-Chromodynamics (QCD) theories \citep{PecceiQuinn77a,PecceiQuinn77b}. In recent years, a wide range of experiments have been designed to detect axion particles and to investigate their possible link to dark matter \citep[see e.g.][for a recent overview]{Banerjee19}: these include e.g. resonant cavity experiments at various frequencies (ADMX \citenp{ADMX20}, ORGAN \citenp{ORGAN}) dielectric haloscopes \citep[MADMAX][]{MADMAX17}, detection-induced magnetic flux oscillations \citep[ABRACADABRA][]{ABRACADABRA19} and NMR-based techniques (ARIADNE \citenp{ARIADNE} and CASPEr \citenp{CASPER}).

The concept of a pseudo-scalar bosonic particle can be generalized from the QCD axion, which is tightly related to the CP problem, to a much broader category of axion-like particles (ALPs) potentially representing dark matter, spanning over an astonishingly wide range of masses of the order $10^{-24} - 10^{0}$ eV$/c^2$ \citep[see][for a comprehensive review on the subject]{Ferreira20review}. The typical wave-like dynamics of the axion acts as an effective net repulsive force, thus admitting a non-degerate self-gravitating stable solution --~called \textit{soliton}~--, whose properties scale with the ALP mass \citep[see e.g.][]{Marsh16review}.

Although --~generally speaking~-- all ALPs share the same proper dynamics, the specific ALP mass sets the cosmological epoch at which the associated dark matter component exits from the oscillatory regime --~which is a peculiar feature of the axion potential~-- and begins to cluster, thus exhibiting different behaviours when ALPs potential role as dark matter is considered \citep[see e.g.][]{Sikivie08}. In particular, a crucial distinction concerns the relative timing of the end of the oscillatory regime with respect to the time of matter-radiation equality: for example, ALPs in the $10^{-10} - 10^{0} \ \rm{eV}/c^2$ range begin to cluster before this time, thus effectively segregating a large fraction of the total dark matter content in gravitationally bound \textit{axion miniclusters}
by the time of baryon decoupling from radiation \citep[see e.g. the early works of][]{Kolb93,Kolb94}.
On the contrary, lighter ALPs density distribution at matter-radiation equality can be essentially described by adding a small-scale correction --~related to its wave-like interaction~-- to the usual CDM density distribution \citep[][]{Hu00}. In this sense, the ALPs dark matter translates in very different cosmological histories whether larger or smaller masses are considered.

\bigskip

In this work, we focus on the lower end of the ALPs mass spectrum --~in the range of $10^{-24}-10^{-19} {\rm eV}/c^2$~-- whose associated dark matter models are often referred to as Fuzzy Dark Matter (FDM). In FDM, the wave-like interaction acts as an net repulsive force and modifies the standard matter power spectrum of CDM at matter-radiation equality, effectively smoothing out density perturbations at small scales and thus leading to fewer collapsed structures at lower redshifts \citep{Hu00}. Moreover, the particle mass is so light that the associated De Broglie wavelength and --~as a direct consequence~-- the self-gravitating objects that can be formed are comparable with the galactic scales \citep[see again][]{Hu00}. These features are of particular cosmological interest, since FDM would simultaneously help solving the putative small-scales inconsistencies of the cusp-core problem \citep{Oh11} and the missing satellite problem \citep{Klypin_etal_1999}.

Numerical simulations of structure formation within FDM models have been initially performed by means of highly numerically intensive Adaptive Mesh Refinement (AMR) algorithms able to solve the Schr\"odinger-Poisson equations over a grid \citep[see e.g.][]{GAMER,GAMER2,Mocz17}, leading to impressive and very detailed results on the properties of individual FDM collapsed objects \citep[see e.g.][]{Woo09,Schive14,Veltmaat18}. However, the computational cost of such approach hindered the possibility to extend the investigation of late time structure formation to large cosmological volumes. To address this issue, N-Body codes were developed, initially only including the (linear) suppression in the initial conditions but neglecting the integrated effect of the FDM interaction during the subsequent dynamical evolution \citep[see e.g.][]{Schive16,Irsic17,Armengaud17} --~i.e. basically treating FDM as standard dark matter with a suppressed primordial power spectrum, similarly to what is routinely done in Warm Dark Matter simulations \citep[][]{Bode00}~--.

In order to exploit the numerical advantages of a N-body approach while not sacrificing the detailing of the FDM dynamics --~crucial in the process of soliton formation~-- throughout the cosmological evolution, the \AG code was developed in \citet{Nori18}. The latter is a modified version of the N-body hydrodynamical cosmological code \PG \citep{Springel05}, that includes the peculiar FDM dynamics through Smoothed Particle Hydrodynamics (SPH) numerical methods, following the approach first proposed in \citet{Mocz15}. The use of SPH techniques to solve for the FDM quantum interactions results in a less numerically demanding algorithm with respect to full-wave AMR solvers, without compromising cosmological results. Therefore, it is now possible with the use of \AG to scale up the volume of FDM simulations related to structure formation and clustering from individual objects to cosmologically representative portions of the Universe \citep[see e.g.][for a list of numerical algorithms used to describe FDM, divided by redshift and scale of interest]{Lague20}.

\bigskip

In this manuscript, which is the first in a series devoted to the study of the scaling relations that characterise the properties of FDM collapsed objects, we present the results obtained in two sets of simulations performed with the \AG code. The first set consists in two sequential simulations of a single collapsing object, aiming to asses the ability of  \AG  to reproduce the typical soliton solution of FDM dynamics in the inner regions of dark matter structures. The second set is composed by high-resolution zoom-in simulations of eight objects extracted from a representative cosmological volume that we use to study in detail the properties of the systems and the scaling relations they exhibit, by allowing them to evolve within their native cosmological context. 

The zoom-in approach consists in a rationalised distribution of resolution elements within the simulation box, which allows to detail a region of interest --~normally, a collapsed structure~-- with high-resolution while efficiently keeping track of its environment \citep[see e.g.][]{Navarro94,Katz94}. In this sense, zoom-in simulations represent an intermediate step bridging single-object simulations and bigger fixed-resolution cosmological simulations. We will proceed towards even larger volumes and thoroughly investigate the possible impact of complex structure formation interactions on scaling relations in the following entry of the series \PARTII.

\bigskip

The paper is organised as follows: in Section~\ref{sec:theory} we briefly describe the FDM models under consideration, providing all the basic equations that enter our numerical implementation (\ref{sec:fdm_th}), and review the scaling relations previously found in the literature that characterise the properties of FDM collapsed objects (\ref{sec:fdm_sr}). In Section~\ref{sec:NM}, we then recall how FDM dynamics is implemented in the \AG code (\ref{sec:AG}), present how collapsed objects are identified (\ref{sec:subfind}) and their related observables are then extracted and computed from the simulation (\ref{sec:properties}). We present and describe the different simulation sets in Sec.~\ref{sec:sims} --~in particular, the collapse of a single object (\ref{sec:single}) and the zoom-in simulations (\ref{sec:zoom})~--. The results are collected in Section~\ref{sec:results}, again presented for the single object (\ref{sec:single_res}) and the zoom-in simulations (\ref{sec:zoom_res}). Finally, in Section~\ref{sec:conclusions} we draw our conclusions.

\section{Theory}
\label{sec:theory}

In this Section, we review the dynamical laws that characterise FDM models, with a special attention to the scaling properties of FDM collapsed structures that arise from the symmetries of the equations and from other assumptions on their morphology and dynamical state.

\subsection{Fuzzy Dark Matter models}
\label{sec:fdm_th}

As we mentioned above, in FDM models the dark matter particle is extremely light, so that the dynamical treatment of dark matter has to take into account quantum interactions. For this reason, FDM is usually described through a quantum bosonic field $\field$, in the assumption of condensation \citep{Hu00,Hui16}.

A massive bosonic field $\field$ evolves according to the Gross-Pitaevskii-Poisson equation \citep[][]{Gross61,Pitaevskii61}
\begin{equation}
\label{eq:GPP}
i \frac{\hslash}{m_\chi} \ \de{t} \field = - \frac {\hslash^2} {m_{\chi}^2} \nabla^2 \field + \Phi \field 
\end{equation}
where $\Phi$ is the Newtonian gravitational potential 
 and $m_{\chi}$ represents the typical mass of the FDM particle.

With the use of the Madelung transformation \citep[][]{Madelung27}
\begin{gather}
\rho = \left| \field \right|^2 \\
\vec v = \frac {\hslash} {m_\chi} \Im{ \frac {\vec \nabla \field} {\field} }
\end{gather}
it is possible recast the problem into a mathematically equivalent fluid description, mapping the field amplitude and phase into a fluid density $\rho$ and a fluid velocity $\vec v$, respectively. In the frame of an expanding universe --~with $a$ and $H=\dot a / a$ being the usual cosmological scale factor and Hubble function, respectively~--, we refer to $\vec x$ as the comoving distance and to the velocity $\vec u$ as the comoving equivalent of $\vec v$. The real and imaginary parts of Eq.~\ref{eq:GPP} then translate into a continuity equation
\begin{equation}
\label{eq:continuity}
\dot \rho + 3 H \rho + \vec \nabla \cdot \left( \rho \vec u \right) =0
\end{equation}
and a modified Euler equation
\begin{equation}
\label{eq:NS}
\dot {\vec u} + 2H \vec u + \left( \vec u \cdot \vec \nabla \right) \vec u =  - \frac {\vec \nabla \Phi} {a^2} + \frac {\vec \nabla Q} {a^4}
\end{equation}
where an additional potential $Q$ --~accounting for the wave-like behaviour of the field~-- appears alongside the usual gravitational potential $\Phi$.  

The gravitational potential $\Phi$ satisfies the standard Poisson equation 
\begin{equation}
\label{eq:poisson}
\nabla^2 \Phi = 4 \pi G \rho_b \ \delta / a
\end{equation}
where $\delta=(\rho - \rho_b)/\rho_b$ is the comoving density contrast with respect to the comoving background density $\rho_b$ \citep{Peebles80}.

The so-called Quantum Potential $Q$ (QP hereafter) has the form
\begin{equation}
\label{eq:QP}
Q = \frac {\hslash^2}{2m_{\chi}^2} \frac{\nabla^2 \sqrt{\rho}}{\sqrt{\rho}}  = \frac {\hslash^2}{2m_{\chi}^2} \left( \frac {\nabla^2 \rho} {2 \rho} - \frac {| \vec \nabla \rho|^2}{4 \rho^2} \right)
\end{equation}
and accounts for the purely quantum behaviour of the field \citep{Bohm52}. It is interesting to remark that, from a theoretical point of view, the QP is present in the usual Euler equation used to describe CDM in cosmology as well: however, it is just safely negligible in the classical limit, as the factor $\hslash^2/m_{\chi}^2$ is extremely small for the typical mass range that has been historically considered for the CDM particle \citep[see e.g.][]{Bertone_Hooper_Silk_2005,Feng10}.

\subsection{Fuzzy Dark Matter: scaling relations}
\label{sec:fdm_sr}

The Euler-Poisson (EP) system composed by Eq.~\ref{eq:NS} and Eq.~\ref{eq:poisson} that governs self-gravitating FDM dynamics reads
\begin{equation}
\label{eq:EP}
\begin{dcases}
\dot {\vec u} + 2H \vec u + \left( \vec u \cdot \vec \nabla \right) \vec u =  - \dfrac {\vec \nabla \Phi} {a^2}+ \dfrac {\vec \nabla Q} {a^4} \\
\nabla^2 \Phi = 4 \pi G \rho_b \ \delta / a
\end{dcases}
\end{equation}
and it admits a spherically symmetric stable solution $\rho_{\rm sol}(r)$ --~usually referred to as the \textit{solitonic core}, since its  density profile is shown to saturate to a constant value in the central regions~--, that has no analytical form but can be well approximated \citep[see e.g.][]{Schive14} by
\begin{equation}
\label{eq:soliton}
\rho_{\rm sol}(r, \rho_c, R_c) = \rhoc \left[ 1 + \alpha \left( \frac{r}{\Rc} \right)^2 \right]^{-8}
\end{equation}
 where $\rhoc$ is the core density
and
\begin{equation}
\label{eq:Rc}
\Rc : \rho_{\rm sol}(\Rc) = \rhoc / 2
\end{equation}
is the half-density comoving radius, simply referred to as core radius, that sets the constant $\alpha  = \sqrt[8]{2}-1$ by construction. In the literature \citep{Schive14}, the core mass $\Mc$ and the soliton total mass $M_{\rm sol}$ have been defined as
\begin{align}
\label{eq:mass_soliton}
\Mc = & \ 4 \pi \int_0^{\Rc} r^2 \rho_{\rm sol}(r) dr \simeq 4 \pi \ (0.2225) \  \rhoc \Rc^3 \\
M_{\rm sol} = & \ 4 \pi \int_0^\infty r^2 \rho_{\rm sol}(r) dr \simeq 4 \pi \  (0.9296) \  \rhoc \Rc^3
\end{align}
where the two quantities are roughly related by $M_c \sim M_{\rm sol} / 4$ due to the different extremes of integration.

In cosmological terms, the net repulsive interaction typical of FDM dynamics results in the presence of a solitonic core in the innermost regions of dark matter structures while recovering the usual CDM behaviour in the outskirts --~as e.g. the typical Navarro-Frenk-White (NFW) density profile~--, where the QP effects are negligible with respect to the gravitational pull.

\bigskip

The EP system of Eq.~\ref{eq:EP} is invariant under the coordinate transformation  via a generic constant $\lambda$ \citep{Ji94}
\begin{equation}
\begin{split}
\label{eq:transformation}
\{ \vec{x}, t,& \vec{u}, \rho, M, \Phi, E \} \\
&\rightarrow \left\{ \lambda \tilde{\vec{x}}, \lambda^2 \tilde{t}, \lambda^{-1} \tilde{\vec{u}}, \lambda^{-4} \tilde{\rho}, \lambda^{-1} \tilde{M}, \lambda^{-2} \tilde{\Phi} , \lambda^{-3} \tilde{E} \right\}
\end{split}
\end{equation}
where we also included the mass $M$ and the energy $E$ of the system. For a detailed treatment of this transformation including the scale factor and the boson mass, see App.~\ref{app:scalefactor}.

\bigskip

It is possible to see that such transformation sets some scaling relations, in particular the core density $\rhoc$, its radius $\Rc$ and its mass $\Mc$ are thus linked through
\begin{equation}
\label{eq:scaling1}
\Rc \propto \left( a \ m_{\chi}^2 \ \rhoc \right)^{-1/4}
\end{equation}
and --~using Eq.~\ref{eq:mass_soliton}~--
\begin{equation}
\label{eq:scaling1bis}
\Rc \propto \left( a \ m_{\chi}^2 \ \Mc \right)^{-1}
\end{equation}
thanks to the intrinsically symmetric nature of the system \citep[see e.g.][for a thorough analytical and numerical study]{Chavanis11a,Chavanis11b}. 

These scaling relations were first explicitly investigated in an astrophysical scenario with dedicated numerical simulations by \citet{Schive14}, where they were confirmed to hold for a sample of haloes at different redshifts in the mass range $10^{9}-10^{11} \dimM$, simulated by directly solving the Schr\"{o}dinger equation on a three-dimensional grid.

In the same work, another important scaling relation linking the features of the soliton core to the properties of the host halo was noticed, namely a relation between the the core mass $M_c$ as defined in Eq.~\ref{eq:mass_soliton} and the virial mass of the halo $M_{vir}$. In this work \citep[as in][]{Schive14}, we use the definition of the virial mass $M_{vir}$, radius $R_{vir}$ and density $\rho_{vir}$
\begin{equation}
\label{eq:virial}
\Mvir = \frac 4 3 \pi \Rvir^3 \ \rhovir = \frac 4 3 \pi \Rvir^3 \ \zeta(a) \ \rho_b
\end{equation}
related to the over-density parameter $\zeta(a)$ as in \citet{Bryan97}.

Linking the core mass to the virial mass is particularly valuable as it allows to estimate properties of the solitonic core for an arbitrary sample of dark matter haloes, based on structural halo properties that can be easily computed e.g. in large-volume simulations of structure formation \citep[as done in e.g.][]{Desjacques19}.
\ \\

However, in order to safely predict core properties with the use of this $M_{c}-M_{vir}$ empirical relation, it is important to review the theoretical assumptions proposed by \citet{Schive14} to justify such relation, as well as the particular conditions of the simulation setup in which this scaling relation was observed.

In fact, this new scaling relation can be heuristically derived by making two strong assumptions regarding the core and halo dynamical states:

\textit{i)} first, the host halo is considered to be in a virialised state in order to be allowed to make use --~in the derivation~-- of the well-known scaling 
\begin{equation}
\label{eq:sigma}
\sigmavir \propto \left(\frac{ \Mvir}{\Rvir} \right)^{1/2} \propto \left( \Mvir \sqrt{\rhovir} \right)^{1/3}
\end{equation}
between the virial mass of a halo and its virial velocity dispersion $\sigmavir$.

\textit{ii)} second, the velocity dispersion of the core $\sigmac$ --~defined as the velocity dispersion within $R_c$~-- is assumed to be comparable to the virial velocity dispersion $\sigmac \sim \sigmavir$.

\bigskip

As a consequence of these two non-trivial assumptions, the core radius $R_c$ and the velocity dispersion of the halo $\sigmavir$ are related via $ \sigmavir R_c \sim 1$. Moreover, it becomes then possible to use Eq.~\ref{eq:sigma} and Eq.~\ref{eq:scaling1} to derive a scaling relation between the virial mass of the halo and the solitonic mass:
\begin{equation}
\label{eq:scaling2}
\Mc \propto \frac{(\Mvir \sqrt{\rhovir})^{1/3}}{\sqrt{a} \ m_{\chi}} \propto \left(\frac{\Mvir}{a \ m_{\chi}^2 \ \Rvir} \right)^{1/2}
\end{equation}
suggesting that massive haloes host soliton cores with higher masses but with smaller radii with respect to less massive systems --~as from Eq.~\ref{eq:scaling1bis}~--.

In their study, \citet{Schive14} tested this latter scaling relation using a suite of numerically simulated FDM haloes and found it to be valid for cores identified in haloes at different redshifts as well as for an individual simulated core during its evolution. The subtle difference between these two cases is of great importance: the former implies that FDM haloes \textit{statistically} satisfy Eq.\ref{eq:scaling2} --~i.e. averaging on the possible dynamical states of  a variety of haloes at a given redshift~--, while the latter suggests that the scaling relation between haloes and the core they host is verified \textit{individually} throughout their history --~ thereby implying that halo evolution does not alter such scaling~--. However, it is worth to remark that in \citet{Schive14}, due to the numerical restrictions on the simulation box size imposed by the grid-approach, the halo sample that is taken into account --~especially at low redshifts~-- seems to fall short in capturing the highly non-linear processes involved in the interaction between different systems --~as e.g. merger events~-- in a broad cosmological setup, focusing on almost isolated and relaxed systems by construction.\\

The scaling relation between the core mass and the virial mass was also investigated in two following works focused on the mergers of FDM solitonic cores, namely \citet{Schwabe16} and \citet{Du16}. 

In \citet{Schwabe16}, the modification of the properties of solitonic cores during merger events was investigated in a non-cosmological framework, by detailing binary merger processes of synthetically produced FDM haloes with different mass ratios, angular momentum and phase difference, as well as multiple merger events. Given the binary merger $M_c = \beta ( M_{c,1}+ M_{c,2})$ with $M_{c,1}$ and $M_{c,2}$ being the masses of two solitonic cores merging into a single final solitonic core of mass $M_{c}$ and $\beta$ being the parameter describing the mass lost in the process, the authors found that the final core mass $M_c$ depends almost entirely on the mass ratio of the two cores involved in the merger. In particular, the final mass $M_c$ is consistent with a value of $\beta \sim 0.7$ whenever $ 3/7\lesssim M_{c,1} / M_{c,2} \lesssim 7/3$, while mergers with more extreme mass ratios result in the dissolution of the smaller core without any significant impact on the bigger one. The authors additionally suggest that, in order to account for different exponents characterising the scaling relation between $\Mc$ and $\Mvir$, it is more appropriate to generalise it as 
\begin{equation}
\label{eq:schwabe}
\frac{\Mc}{\Mvir} \propto \left( \frac{ \rhovir^{1/3}}{a \ m_{\chi}^2 \ \Mvir^{4/3}} \right)^{\eta} \propto \left( a \ m_{\chi}^2 \ \Mvir \Rvir \right)^{-\eta} 
\end{equation}
such that this generalised scaling relation is equivalent to Eq.~\ref{eq:scaling2} in the case $\eta=1/2$. Based on the relation between the core mass and total mass of the final systems --~where the total mass can be associated with the virial mass~--, the authors were not able to pinpoint a specific scaling due to the large scatter in the the range $\eta \in [1/2, 1/6]$.

\bigskip

Following up on the results of \citet{Schwabe16}, \citet{Du16} account neither for the combined evolution of the core mass and halo virial mass after the initial collapse nor for the role of merger events, and 
proposed a parameterisation of the scaling exponent of Eq~\ref{eq:scaling2} as a function (solely) of the core mass loss parameter $\beta$, based on the estimate of the number of mergers that a typical system undergoes during its evolution --~and, in particular, the mergers with a large enough mass ratio to alter the core mass~--, providing the scaling relation
\begin{equation}
\label{eq:scaling3}
\Mc \propto B(\beta,a,m_{\chi}) \ \Mvir^{\ \log_2(2\beta)}
\end{equation}
where the normalisation factor $B(\beta,a,m_{\chi})$ accounts for an estimated number and type of merger events for a given halo mass distribution, which depends both on redshift and on the parameter $\beta$ \citep[see][for technical details]{Du16}. In their work, the authors expand the exponent $\log_2(2\beta)$ into $(2\beta-1)/\ln(2)$ assuming $\beta \sim 0.5$, but end up using $(2\beta-1)$ for their analysis, in order to impose $\Mc \propto \Mvir$ for $\beta=1$. These various forms clearly generate some confusion, as the value $\beta\sim0.7$ found by \citet{Schwabe16} maps in quite different exponents, namely $\eta \sim 0.386, 0.317, 0.45$ expressed in terms of Eq.~\ref{eq:schwabe} using the first, second and third form, respectively.

\bigskip

An additional suggestion on the subject came from a work of another independent group \citep{Mocz17}, where a larger sample of solitonic core mergers were simulated in a non-cosmological framework. As in \citet{Schwabe16}, the scaling relation of Eq.~\ref{eq:scaling2} was not confirmed and the alternative relation
\begin{equation}
\label{eq:scaling2bis}
\frac{\Mc}{\Mvir} \propto \left( \frac{ \rhovir^{1/3}}{a \ m_{\chi}^2 \ \Mvir^{4/3}} \right)^{1/3} \propto \left( a \ m_{\chi}^2 \ \Mvir \Rvir \right)^{-1/3}
\end{equation}
was observed in its place, equivalent to Eq.~\ref{eq:schwabe} with the exponent $\eta = 1/3$. Based on theoretical considerations, \cite{Mocz17} point out that such relation is retrieved by replacing the close connection between dispersion velocities of the core and the halo ($\sigmac \sim \sigmavir$) that was invoked by \citet{Schive14} by a similar relation involving the core and halo energies $\Mc \sigmac^2 \sim \Mvir \sigmavir^2$.

\bigskip

The challenging task of estimating a universal scaling relation between the core and the halo mass reflects the complexity of the processes involved in halo formation and their impact on the solitonic core properties \citep[see e.g.][for an interesting discussion on such scaling relation]{Bar18}. The mass of the core and the virial mass of the host halo are related to each other due to their co-evolution as parts of the same larger system, yet they individually obey to different dynamics and are in contact with different environments: restricting the analysis only to relaxed systems \citep[as in][]{Schive14,Schwabe16,Mocz17} allows to reduce the impact of such complexity, at the cost of predictability on cosmologically realistic halo populations --~which  necessarily include systems in highly different dynamical states and stages of evolution~--. This task becomes even more daunting if non-linear cosmological time-dependence is taken into account --~as \citet{Schwabe16,Du16,Mocz17} did not~--, since the validity of the approximations introduced in this Section and of the very EP systems of Eq.~\ref{eq:EP} may vary over time.

\bigskip

In order to be as general as possible, in the following we will investigate the relation between the core and halo properties $\left( \Rc, \rhoc, \Mc, \Mvir, \Rvir \right)$ making use of the relations
\begin{gather}
\label{eq:SR}
\Rc = \kappa \ \left( \frac{10^{10} \dimrho} {a \ \rhoc} \right)^{\mu} \left( \frac{2.5}{m_{22}} \right)^{2\mu} \dimR \\
\frac {\Mc}{\Mvir} = \tau \ \left( \frac{10^{10} \dimM}{a \ \Mvir} \right)^{\eta} \left( \frac{\dimR}{ \Rvir} \right)^{\eta} \left( \frac{2.5}{m_{22}} \right)^{2\eta}
\end{gather}
that we will term Scaling Relation I (SRI) and Scaling Relation II (SRII), respectively, where $\kappa$ and $\tau$ are normalisation factors to be estimated along with the $\mu$ and $\eta$ exponents. Here, we used the definition $m_\chi = m_{22} \times10^{-22} \ \text{eV/c}^2$ to parameterise the boson mass. The system of equations is closed by the definition of Eq.~\ref{eq:mass_soliton} that we use to derive the soliton mass from its density and radius. Using Eq.~\ref{eq:virial} it is possible to express $\Rvir$ in terms of $\rhovir$, but we prefer the latter over the former because of the more elegant mathematical form of the resulting equations.

\bigskip

Let us remark once again that the scaling relations SRI and SRII are linked to different and independent assumptions and approximations: the former results from the symmetries of the spherical ground-state solution of the EP system at the core scale, while the latter invokes the analogy between the core and the halo velocity dispersions and the virialisation of the host system, thus implying the sphericity of the whole dark matter halo and the relaxed nature of its dynamics. The virialisation assumption is more stringent and may easily imply the sphericity at the core-level, but not \textit{viceversa}. Therefore, in a cosmologically representative volume, it is reasonable to expect particularly un-relaxed FDM systems, in which the core is not yet stabilised in its ground-state solution --~i.e. it is not yet a proper \textit{solitonic} core~--, to be inconsistent both with SRI and SRII, while other haloes, harbouring spherically symmetric ground-state solitonic cores, to satisfy SRI --~satisfying or not SRII, depending on their global virialisation state~--. Hence, for a proper determination of global scaling relations (and their associated scatter) between the structural properties of the solitonic cores and those of their host haloes, it is of great importance to investigate thoroughly the dynamical state of the hosts to better discern cases in which scaling relations should hold from cases where deviations are expected. 

\section{Numerical Methods}
\label{sec:NM}

In this Section, we brefly review the relevant properties of the \AG code \citep{Nori18} and the halo-finding algorithm \SUBFIND \citep{Springel_etal_2001} that we employ to run and analyse the simulations discussed in this work. We then define the main observables of interest regarding collapsed objects and present the numerical methods we use to extract them from simulations.

\subsection{The code: \AG}
\label{sec:AG}

\AG \citep[presented in][]{Nori18} is a module  of the cosmological and hydrodynamical N-Body code \PG  \citep[a non-public extension of the public {\small GADGET}2 code, ][]{Springel05}, that implements the physics of FDM models in cosmological simulations of structure formation.

Following the N-body approach of \PG, the density field and its derivatives --~and, ultimately, accelarations~-- are reconstructed from the distribution of discrete tracers --~i.e. particles~-- via refined Smoothed Particle Hydrodynamics (SPH) routines. For any technical detail that goes beyond the short description provided below, we refer the reader to \citet[][]{Nori18}. 

The general SPH approach relies on the concept that a continuous observable $O$ that underlies a discrete set of fluid-element particles can be approximated at particle $i$ position with the sum on neighbouring particles $j \in \text{NN}(i)$ --~which includes particle $i$ itself~-- weighted on particle mass $m$ and a kernel function $W_{ij}$ of choice.
Such approximation reads
\begin{equation}
O_i = \sumnn m_j \frac {O_j} {\rho_j} W_{ij}
\end{equation}
and can be extended to spatial derivatives as
\begin{equation}
\nabla O_i = \sumnn m_j \frac {O_j - O_i} {\sqrt{\rho_j\rho_i}} \nabla W_{ij}
\end{equation}
where $\rho$ is the density field, which can be calculated as
\begin{gather}
\rho_i = \sumnn m_j W_{ij},
\end{gather}
with the very same approach by taking $O \equiv \rho$.

The kernel function $W_{ij}$ has the physical dimension of an inverse volume and heuristically represents the probability of finding particle $i$ at position $r=|\vec r_i - \vec r_j|$. The typical measure of this uncertainty volume is given in terms of the so-called smoothing length $h_i$, whose extent is fixed by imposing
\begin{equation}
\label{eq:NN}
\frac{4}{3} \pi h_i^3 \rho_i= \sumnn m_j
\end{equation}
which is equivalent to fixing the mass enclosed within $r\leq h_i$. \\

The complete scheme used by \AG to reconstruct the particle acceleration due to the QP of Eq.~\ref{eq:QP} is then based on the same SPH general approach, reading:
\begin{gather}
\vec \nabla Q_i =  \frac {\hslash^2}{2m_{\chi}^2} \sumnn \frac {m_j} {f_j \rho_j} \vec \nabla W_{ij} 
\left( \frac {\nabla^2 \rho_j} {2 \rho_j} - \frac {| \vec \nabla \rho_j|^2}{4 \rho_j^2} \right)
\end{gather}
where $f$ is a factor that accounts for the adaptive adjustment of the smoothing lengths of each single particle \citep[see][for details]{Nori18}.

\bigskip

In \PG, and \AG as well, N-body particles are divided up to six different \textit{types} that are meant to represent fluids characterised by different dynamics: the historical reason behind this implementation is to numerically differentiate the collisionless dark matter species from the hydrodynamical one representing baryonic gas, as well as from the collisionless population of stars and black holes that may include all possible non-linear radiative processes. Since in \AG we introduce FDM particles as a separate type, we can perform simulations of a mix of dark matter particles following CDM and FDM dynamcs: this feature will be used in Sec.~\ref{sec:zoom} to approximate the behaviour of FDM to the one of CDM at the largest scales.

\AG has undergone various stability tests and has proven to be not only less numerically intensive with respect to Adaptive Mesh Refinement (AMR) full-wave solvers \citep{GAMER}, due to the intrinsic SPH local approximation, but also to be accurate for cosmologically relevant scales as it agrees both with the linear \citep{axionCAMB} and the non-linear results \citep{Woo09} available in the literature, even if a proper convergence and code comparison test among the various different implementations of FDM that have been developed in the literature has not yet been performed, which would be necessary to assess the consistency of different numerical methods at very small scales.

\subsection{Halo finder and merger tree construction}
\label{sec:subfind}

To identify collapsed structures in our simulations we used the \SUBFIND Friends-Of-Friends (FoF) algorithm \citep[][]{Davis_etal_1985} with an unbinding procedure to identify gravitationally bound substructures within the FoF ensembles \citep{Springel_etal_2001}.

The unbinding procedure is based on energy balancing given by the virial theorem, that in the FDM framework includes the effects of QP \citep{Hui16} as well as the ones of kinetic energy $K$ and gravitational potential $\Phi$, extending $2K + \Phi = 0$ to $2K + \Phi + 2Q = 0$ with respect to the standard CDM case.

Since our aim is to investigate the properties of solitonic cores --~which satisfy by construction the \textit{quantum} virial theorem \citep{Hui16}~-- and describe them in terms of general global properties, we decided to use \SUBFIND with the standard virial theorem for simplicity. We take care of noting that, given the net repulsive effects of the QP on small scales, the use of the quantum version of the virial theorem would have the following consequence: the particles in the halo outskirts that are found to be weakly bound could instead not bound at all to the main structure. Even though this does not impact on the properties of the solitonic core when large systems are considered, the smallest collapsed structures found in the simulations --~described by a small number of particles~-- could be \textit{de facto} unbounded. For these reasons and to avoid numerical artefacts related to poorly resolved systems, it would be then advisable to discard haloes that are described by a small amount of particles and restrict the analysis on halo properties that are statistically insensitive to particles in the outskirts --~as suggested for the $\Lambda$CDM case e.g. in \citet{Neto_etal_2007}, by imposing a minimum threshold of $1000$ total particles per halo~--. In our case, the zoom-in simulated haloes that will be discussed below have all more than $10^{5}$ particles and have central densities that are three to four orders of magnitude greater than the mean matter density at redshift $z=0$, so we can safely neglect the contribution of the QP in the unbinding process for the estimation of core features.

Hereafter, we use the terms \textit{primary structures} to identify the substructures of each FoF group containing the most gravitationally bound particle, \textit{subhaloes} for the non-primary structures and \textit{haloes} when we generally consider the whole sample of structures --~i.e. \textit{primary} and \textit{non-primary}~-- identified by our halo finging procedure.

\bigskip

Since we are interested in the evolution of halo properties in time, we will also need a procedure to link haloes across different redshifts, by identifying connections between any given halo and its progenitors/descendants in order to understand its formation history. To this end, the halo catalogues were combined to form merger trees, using the methods and definitions described in \citet{Springel05,Aquarius}.

Given the hierarchical evolution of cosmological structures, the reconstruction of mergers trees consists in identifying a common share of particles within each halo at lower redshifts with the ones of a halo --~or more haloes, in the case of merger events~-- at higher redshifts. To this end, N-body particles are flagged by a fixed and unique ID throughout the simulation that is used in the identification process. 

The accuracy of this reconstruction is clearly bounded by the time resolution of the finite set of redshifts $\{z_n\}$ at which outputs are produced. Ideally, one would produce as many outputs as possible to maximise the time resolution, however an extreme time resolution can be redundant --~i.e. if the redshift spacing is very small, outputs would be almost identical to each other with a large majority capturing no merger events~-- and quickly leads to exceeding the available amount of memory storage. Our set of output redshifts is such that in the interval $0 \leq z \leq 2$ an output is produced every $\Delta z = 0.2$, with the addition of outputs at higher redshifts $z \in \{2.33,3,4,5,9\}$, which allows for a good time resolution and no redundancy.

\subsection{Halo properties}
\label{sec:properties}

In this Section, we list the physical observables that are relevant to our analysis and describe the strategies we used to compute them from the available simulation outputs.

\subsubsection{Sphericity and Centre Offset}

The shape of a halo can be a useful observable to understand its dynamical state and has been shown to correlate with dynamical features \citep[see e.g.][and references therein]{Neto_etal_2007, Maccio_etal_2008}.

To define the shape of haloes, we use the inertia tensor of the halo member particle ensemble as identified by \SUBFIND to be gravitationally bound:
\begin{equation}
I_{ij} = \sum_{particles} m\ (\hat e_i \cdot \hat e_j)\ |r|^2 - (\vec r \cdot \hat e_i)\ (\vec r \cdot \hat e_j)
\end{equation}
where $\hat e$ are the unit vectors of the reference orthonormal base and $\vec r$ and $m$ are the particle position and mass, respectively. The equivalent triaxial ellipsoid with uniform mass distribution can be built from the eigenvalues and the eigenvectors of the tensor, each representing the square moduli and unitary vectors of the main axes. We define $a \geq b \geq c$ the lengths of the three axes and the sphericity $s = c/a$. Note that $s$ represents the sphericity of the total system comprehensive of the core and the host halo: it would be tricky to define two separate sphericities for the core and the halo, also because our identification procedure for the cores is based on the typical density profile of Eq.~\ref{eq:soliton} that assumes sphericity at the core level.

Clearly, the calculation of the inertia tensor and, therefore, the sphericity of a halo depends on the choice of the centre of the system. The center of a dark matter halo in N-body simulations is usually identified either by the position of the most bounded particle $\vec r_{\rm MB}$ (i.e. the particle closest to the local minimum of the halo gravitational potential) or by the centre of mass of the system $\vec r_{\rm CM}$. These two definitions are statistically equivalent for relaxed and isotropic systems, where the particle spatial distribution and the gravitational potential are consistent. Conversely, unrelaxed and small systems may show deviations between the two definitions, especially in particularly anisotropic environments where the neighbouring systems effect on the underlying gravitational potential is not negligible, as e.g. for subhaloes. In fact, we will use the centre offset
\begin{equation}
\label{eq:offset}
d_{\rm off}= | \vec r_{\rm MB} - \vec r_{\rm CM} |
\end{equation}
to be coupled with the sphericity as a proxy for dynamical relaxation \citep[as e.g. done in][]{Neto_etal_2007}.

\subsubsection{Density profile and soliton fit}
\label{sec:DP}

To estimate the various observables related to the radial density profile of haloes as the virial mass $\Mvir$ and the virial radius $\Rvir$ of the halo as well as the density $\rhoc$ and radius $\Rc$ of the core, we need to define a consistent numerical procedure to build the halo density profiles. In our analysis, we found that different strategies used to compute the density profiles may have a non-negligible impact on these quantities, especially on the estimate of $\rhoc$ and $\Rc$ that are linked to the inner region of the radial profile. In particular,
the choice of the halo centre (discussed in the previous Subsection) and the numerical evaluation of the density in each individual radial bin both play an important role.

A straightforward approach to calculate the density profile $\rho(r)$ consists in counting the N-body particles in each of a set of radial bins --~corresponding to a spherical shell~-- and take the ratio between the total particle mass in the bin and the bin volume. Due to the radial nature of the observable and the discreteness of N-body simulations, however, the estimate of the density in the innermost regions carries a great statistical error, thus hindering a good evaluation of $\rhoc$ and the solitonic profile.

Thanks to the \AG design, we can instead rely on the particle density as computed by the SPH routine, hence greatly reducing the errors especially related to low particle counting in the regions we are most interested in. The density profiles of haloes are thus computed as the mean SPH density of particles in each spherical shell. The quantities $\rhoc$ and $\Rc$ are fitted independently on the profiles by a two parameter logarithmic fit, based on Eq.~\ref{eq:soliton}, while $M_{vir}$ and $r_{vir}$ are estimated as from Eq.~\ref{eq:virial}.

\subsubsection{Formation time and mass gained via merger}
\label{sec:mergertree}

From our merger trees, we are able to extract very important information about the formation and evolution of structures, in particular, the approximate formation time of the halo and the mass that has been accreted through mergers. As previously mentioned in Sec.~\ref{sec:subfind}, the accuracy of these quantities is bounded by the time resolution of the simulations snapshots, as they are restricted to the finite set of available redshifts $\{z_n\}$; however, they provide very useful insights on the evolution of a system, which will be particularly relevant for our discussion.

For each halo $i$ of interest at a given time $z_n$, we identify the haloes $P(i)$ found at $z_{n-1} > z_n$ that merged into halo $i$ as its \textit{progenitors}, and the one halo 
$D(i)$ at $z_{n+1} < z_n$ whose progenitor is the halo $i$ as its \textit{descendant}. We refer to the sequence $L(i)$ of $i$ and its most massive progenitors at each redshift $\forall z \geq z_n$ as \textit{direct line} and we define the redshift of formation $z_{\rm form}$ as the highest redshift the direct line extends to.

To estimate the impact of mergers on the halo properties in time, we define the mass $M^i_{\rm merg}$ of halo $i$ as the cumulative mass of the progenitors of the haloes belonging to the direct line $L(i)$ that do not belong to the direct line themselves, i.e.
\begin{equation}
M^i_{\rm merg}(z_n) = \sum_{j \in L(i)} \ \sum_{p \in P(j) \ \textbackslash \ L(i)} M_p
\end{equation}
where $z_n$ and $z_m$ are part of the discrete set of redshifts. In this way, $M^{i}_{\rm merg}(z_n)$ physically represents the mass share that halo $i$ has cumulatively gained via mergers during its history, from its formation up to $z_n$. As an additional condition, in the calculation of $M_{\rm merg}$ we only take into account contributions related to merger events characterised by a mass ratio of $1:20$ or higher, in order for $M_{\rm merg}$ to be safely independent from resolution.

\begin{figure}
\includegraphics[width=\columnwidth,trim={1.4cm 1.4cm 3.5cm 3cm},clip]{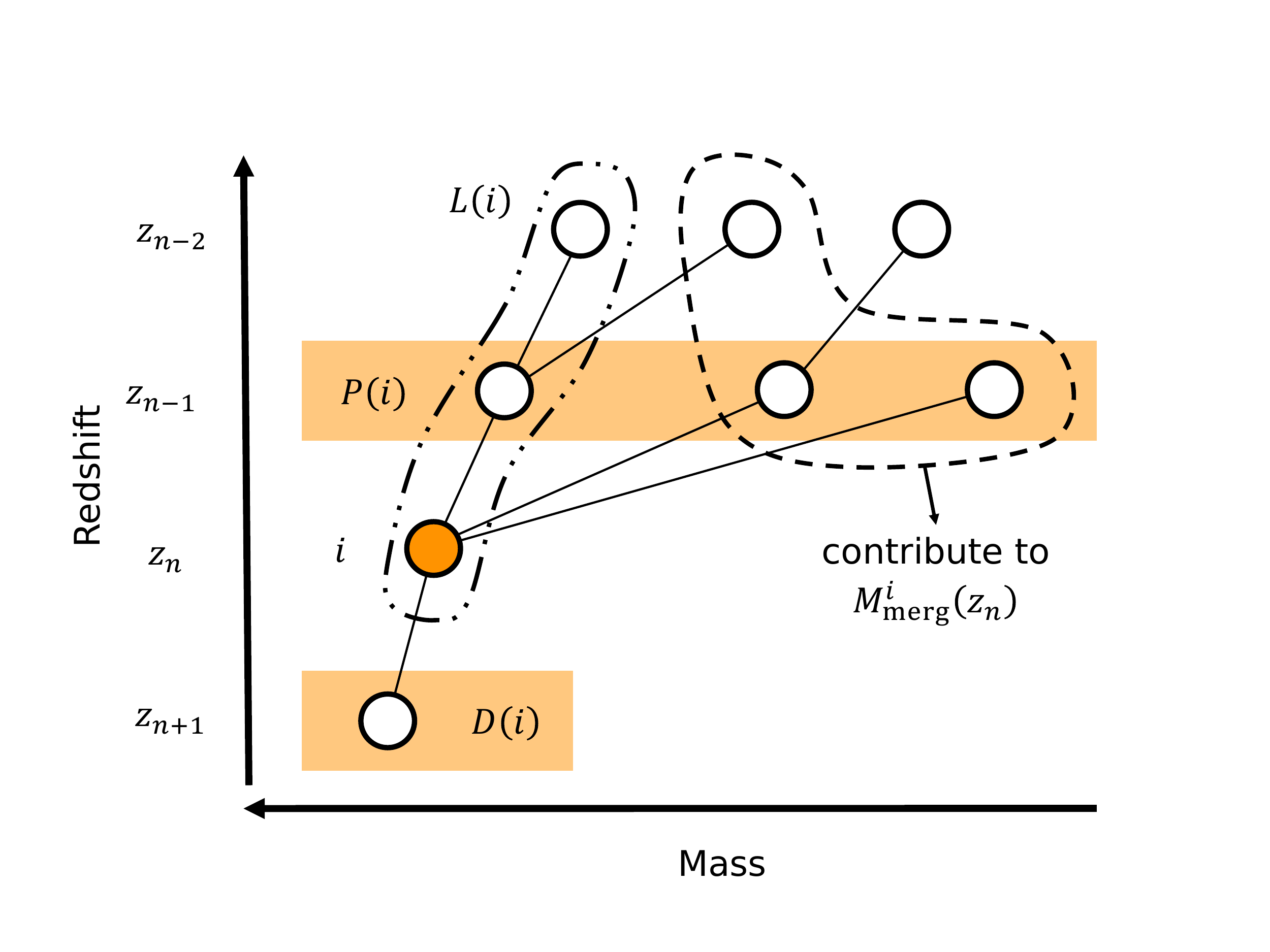}
\caption{Schematic representation of haloes (circles) belonging to a single merger tree in the mass-redshift plane, where we visually display the ensembles and quantities defined in Section~\ref{sec:mergertree} for a given halo $i$ at redshift $z_n$ (highlighted in orange).
}
\label{fig:merger_tree}
\end{figure}

\section{Simulations}
\label{sec:sims}

In this Section, we introduce and describe the two simulation setups that will be presented in this work: the first one focuses on the collapse of a single object, to assess the ability of \AG to correctly reproduce a solitonic core at the centre of haloes; the other one is a set of \textit{zoom-in} simulations of individual haloes --~extracted from a parent low-resolution cosmological run~-- aiming to study the properties of FDM haloes and solitonic cores as \textit{individual} systems in a cosmologically realistic environment. For all the simulations, we assume the totality of matter to be composed by FDM with a particle mass of $m_{22} = 2.5$. The cosmological background parameters used for all the simulations are $\Omega_m = 0.317$, $\Omega_{\Lambda} = 0.683$, $H_0 = 67.27 \ \text{km/s/Mpc}$ together with the initial power-spectrum parameters $n_s = 0.965$ and $\sigma_8 = 0.816$. To build the cosmological initial conditions of our simulations, we used the code \MUSIC \citep{music} together with the \AC \citep{axionCAMB} solver to compute the correct suppressed matter power spectrum at the initial redshift.

\subsection{Single object collapse}
\label{sec:single}

The ability of \AG to correctly reproduce the typical solitonic core feature in the innermost regions of dark matter haloes was supported by the result of an idealised test \citep[presented in][]{Nori18}, but it was not possible to confirm it in a realistic cosmological context, since the spatial resolution of the previous works involving \AG applications was comparable with the typical scale of core size \citep[][]{Nori19}.

To assess whether \AG is actually able to reproduce proper solitonic cores, we performed a simple yet meaningful test: we simulated a cosmological box of side $L = 500\, \, \dimR$ with $128^3$ particles, corresponding to a mass resolution and softening length of $5.244 \times 10^3\, \,  \dimM$ and $100$  pc$/h$ respectively, and let it evolve with periodic boundary conditions from redshift $z=99$ to $z=0$. The box size was chosen to contain approximately one (quantum) Jeans mass, in order to form a single object for which boundary conditions ensure no loss of mass and energy.

Upon reaching $z=0$, we extended the simulation by switching off the expansion of the universe --~i.e. effectively keeping $z=0$ constant~-- to investigate the stability of the system independently of redshift.

Although starting from cosmological initial conditions, this test has numerical and physical yet limited cosmological value, since it is a highly idealised system primarily focused on the ability of the \AG code to reproduce a stable solitonic core within a dark matter halo; to investigate cosmologically relevant systems we resort to \textit{zoom-in} simulations described in the next Section.

\subsection{\textit{Zoom-in} simulations}
\label{sec:zoom}

In order to reconstruct in detail the solitonic core structure that forms within FDM haloes, we performed a series of \textit{zoom-in} cosmological simulations \citep[see e.g.][]{Navarro94,Katz94}. The zoom-in approach allows to reach high resolution in a selected region of the simulation while still following the evolution of the surrounding cosmological environment with a coarser resolution, thus neglecting the fine details outside the region of interest to greatly reduce the simulation run time with respect to a fixed resolution approach. By differentiating the concentration of the resolution elements --~namely N-body particles or cells, depending on the nature of the simulation~-- within the simulation domain, this technique is particularly useful to study single objects in detail without loosing completely information on the cosmological environment, with a great improvement on the computational cost with respect to a fixed-resolution approach \citep[see e.g.][for a review on the subject]{Kuhlen_Vogelsberger_Angulo_2012}.

This method involves three steps: first, a preliminary simulation with low resolution is performed to identify the collapsed structures at low redshift among which the target structures for the zoom-in runs are selected; second, the particles belonging to the selected haloes are mapped back to the initial conditions, to estimate the extent of the original Lagrangian region of each halo; third, the Lagrangian region is re-populated with a larger number of (less massive) particles to characterise the density field with a higher resolution. Usually, a tier of decreasing refinement levels is imposed outside the region of interest to avoid a sharp transition in resolution, with the last level accounting for the farthest regions having a significantly lower resolution than the preliminary uniform simulation.

To implement this procedure we resorted to the public code \MUSIC \citep{music}, that we used in the first place to build the initial conditions for the preliminary simulation --~termed \textit{COARSE}~-- at redshift $z=99$, with $256^3$ particles in a box of side $15 \text{Mpc/h}$, resulting in smoothing length of $\epsilon_\text{res} \sim 1 \text{ Kpc/h}$. The initial power spectrum provided to the \MUSIC code to realise particle displacements was computed with the code \AC \citep{axionCAMB}, that coherently suppresses small-scale power as required in the FDM framework. 

As \AG allows for the QP interaction to be switched on or off --~i.e. evolve particles with a FDM or a standard CDM dynamics, respectively~--, we choose not to include the QP interaction in the dynamics of this preliminary \textit{COARSE} simulation, due to the marginal effects of the QP at the scales, redshifts and masses of interest and this first explorative nature of the simulation \citep[as seen in][]{Nori19}.

\begin{table}
\caption{Technical properties of the preliminary and zoom-in simulations. The minimum and maximum values of resolution levels --~$l_{\text{min}}$ and $l_{\text{max}}$~-- as well as the volume at maximum resolution $V_{\text{max}}$ are given with respect to the COARSE simulation.}
\label{tab:SIMS}
\centering
\begin{tabular}{cccccc}
\hline
\hline
\multirow{2}{*}{Name}   & \multirow{2}{*}{$l_{\text{min}}$} & \multirow{2}{*}{$l_{\text{max}}$} & \multirow{2}{*}{$V_{\text{max}}$} & $M_{\text{res}}$ & $\epsilon_{\text{res}}$ \\
 & & & & $ [\dimM]$ & $[\mathrm{pc}/\mathrm{h}]$ \\
\hline
COARSE & $1$   & $1$    & $100 \%$                 & $1.770\times 10^{7}$      &    $1000$  \\
\hline
A      & $2$   & $1/4$  & $5.89 \%$                & $2.765\times 10^{5}$      &    $250$   \\
B      & $2$   & $1/4$  & $3.98 \%$                & $2.765\times 10^{5}$      &    $250$   \\
C      & $4$   & $1/4$  & $2.16 \%$                & $2.765\times 10^{5}$      &    $250$   \\
D      & $4$   & $1/4$  & $1.41 \%$                & $2.765\times 10^{5}$      &    $250$   \\
E      & $4$   & $1/4$  & $1.01 \%$                & $2.765\times 10^{5}$      &    $250$   \\
F      & $4$   & $1/8$  & $0.65 \%$                & $3.456\times 10^{4}$      &    $125$   \\
G      & $4$   & $1/8$  & $0.10 \%$                & $3.456\times 10^{4}$      &    $125$   \\
H      & $4$   & $1/8$  & $0.13 \%$                & $3.456\times 10^{4}$      &    $125$   \\
\hline
\hline
\end{tabular}
\end{table}

From the structures identified in the \textit{COARSE} simulation at redshift $z=0$, we chose eight haloes to be simulated again with a zoom-in approach, which we label with letters from $A$ to $H$, spanning over more than two orders of magnitude in virial mass.

To avoid contamination with particles of different mass in the central region of the haloes \citep[and in line with previous works, as][]{Neto_etal_2007}, we conservatively extended the region of maximum resolution by $2.5$ times in each direction with respect to the smallest cuboid in the initial conditions that encloses all the particles belonging to the target halo at $z=0$.

Since \MUSIC refines initial conditions using levels in a grid-approach with relative spacing in powers-of-two, we downgraded the resolution of regions outside the domain of interest by a factor $4$ or $2$ with respect to the \textit{COARSE} simulation --~i.e. in N-body terms, we reduced the mean inter-particle distance by using less and more massive particles~--, while the maximum refinement within the high-resolution region reached a factor of $1/4$ or $1/8$ depending on the system. The refinement factors of the low- and high-resolution regions with respect to the \textit{COARSE} simulation --~termed $l_{\rm min}$ and $l_{\rm max}$, respectively~-- and the volume fraction $V_\text{max}$ identifying the region of maximum resolution in the initial conditions are summarised in Tab~\ref{tab:SIMS}, together with the mass $M_\text{res}$ and softening length $\epsilon_\text{res}$ of the smallest resolution elements.

In our work, intermediate levels of refinement devised to smoothly transition between these extremes were assigned to a different particle type of the \PG particle data structure to allow for an easier identification in post-processing analyses and for a different treatment of their dynamics in terms of QP contribution. In fact, in the spirit of performance enhancement of zoom-in simulations, we decided to follow the full FDM dynamics including the effect of the QP only for particles representing the highest level of resolution, while neglecting this contribution for the low-resolution levels. Since the solution of FDM dynamics in \AG relies on matter density and its derivatives as calculated on neighbouring particles, we also included particles representing the second-highest resolution level in the calculation: for these particles, laying just outside the region of interest, we compute the density and its derivatives as discussed in Sec.~\ref{sec:AG} but no QP contribution to acceleration was added, thus behaving as an effective buffer between high- and low-resolution regions that greatly reduce the errors in the dynamics in the outskirts of the high-resolution domain. As a further check of our implementation, we also simulate the same zoom-in systems without including the QP, to have a direct proof of the role of the QP in the formation and stability of the solitonic core in the innermost regions of dark matter haloes. 

In the selection process, we first defined a set of mass bins for the target structures at $z=0$, in order to include a variety of final halo masses in our sample of zoom-in simulations. Then, within each of these mass bins, we preferred --~as a first selection criterion~-- haloes with the smallest initial Lagrangian region. This preference sourced from two main practical considerations: first, for obvious numerical reasons, the smaller the high-resolution volume is, the less computationally intensive the simulation becomes; second, we wanted to avoid haloes undergoing extreme merger events during their evolution, that intuitively map --~for fixed final halo mass~-- into a larger portion of the simulation volume once traced back to the initial conditions. We also preferred --~as a second selection criterion~-- haloes that formed quite early in order to have information on their evolution in a larger number of simulation snapshots. The impact of this selection bias will be addressed in Sec.~\ref{sec:conclusions} were results are discussed.

\bigskip

\section{Results}
\label{sec:results}

In this Section, we present the results obtained in the single object collapse test and in the set of zoom-in simulations. In particular, we first detail the properties of the core obtained in the former test, we then present and discuss the properties of haloes and solitonic cores forming in the latter set of simulations, extracting valuable information on the scaling relations linking different observables as discussed in Sec.~\ref{sec:fdm_sr}, to investigate regimes of agreement and deviation from global trends.

\subsection{Single object collapse}
\label{sec:single_res}

\begin{figure}
\includegraphics[width=\columnwidth,trim={0.5cm 0.5cm 0.5cm 0.5cm},clip] {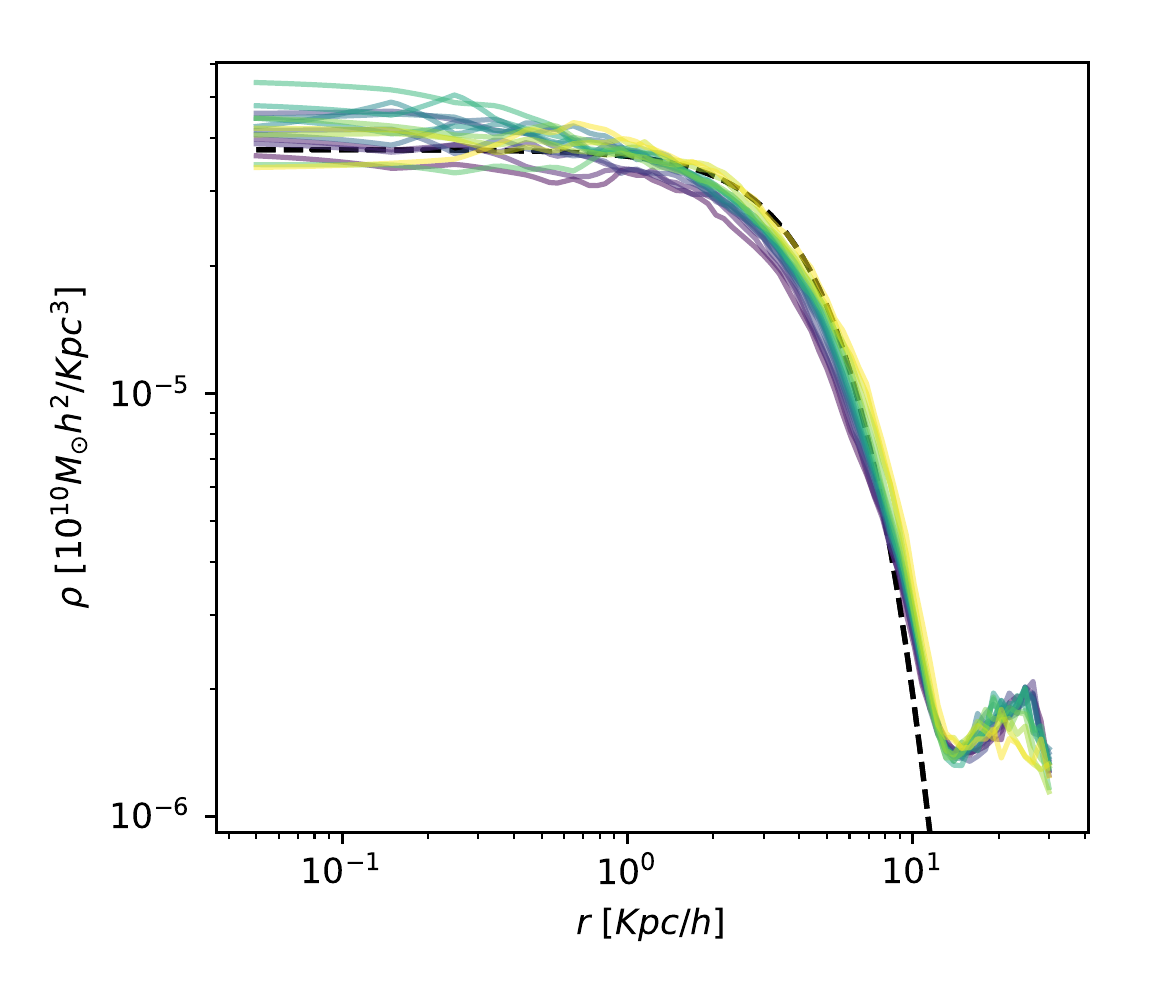}
\caption{Density profile of the halo formed in the single object collapse test, as observed at different times in the non-cosmological extension of its evolution (colored solid lines). The dashed black line represent the analytical profile of the solitonic core as fitted using Eq.~\ref{eq:soliton}.}
\label{fig:single}
\end{figure}

As expected, only one halo forms within the simulation box of the single object collapse test. In Fig.~\ref{fig:single} the radial density profile of the halo at $z=0$ is shown, together with its evolution in the extended non-cosmological part of the simulation. It is possible to see that the halo formed in this test consists almost entirely of the solitonic core with negligible outer features. The density profile is properly described by Eq.~\ref{eq:soliton} represented by a black dashed curve, corresponding to the mean core radius and density observed at the end of the extended simulation beyond $z=0$. The mean properties of the core and the halo are summarised in Tab.~\ref{tab:single}. Let us remark that the virial mass of this object is approximately $\Mvir \sim 4 \Mc \sim M_{\rm sol}$, confirming that what we identify here as \textit{halo} coincides with the whole soliton. 

\begin{table}
    \centering
    \caption{Properties of the solitonic core and the halo formed in the single object collapse test.}
    \begin{tabular}{ccccc}
    \hline
    \hline
    $\Rc$ & $\rhoc$ & $\Mc$ & $\Rvir$ & $\Mvir$ \\
    $[\dimR]$ & $[\dimrho]$ & $[\dimM]$ & $[\dimR]$ & $[\dimM]$ \\
         \hline
         $4.496$ & $3.791\times10^{5}$ & $9.63\times10^{7}$ & $14.301$ & $3.69\times10^8$ \\
         \hline
         \hline
    \end{tabular}
        \label{tab:single}  
\end{table}

\bigskip

For this single halo, it is possible to estimate the normalisation factor $\kappa \sim 0.353$ for SRI and, similarly, the values of $\tau \sim 0.190$ for SRII, by assuming $\mu = 1/4$ and $\eta=1/2$, respectively. Let us remark that the bare value of these normalisation factors does not hold any particular physical meaning \textit{per se}. Nevertheless, since solitons are described by a family of functions which is invariant under the transformation of Eq. \ref{eq:transformation} --~that results in the scaling relation SRI, if the sphericity of the system is assumed~--, the normalisation value should be unique for all the spherically symmetric solitonic cores. We thus expect to find a similar value of $\kappa$ for the solitons identified in our zoom-in simulated haloes (discussed in the next section), at least as long as the spherical approximation is valid. If we also assume SRII to hold, the same uniqueness property would be valid also for $\tau$; however, as discussed in Sec.~\ref{sec:fdm_sr}, SRII is expected to hold only if the dynamical state of the halo-core system is considered to be identical for all systems, which might not to be the case in a realistic cosmological setup.

\subsection{Zoom-in haloes}
\label{sec:zoom_res}

In this Section, we present the properties of all zoom-in simulated haloes, both in terms of their general structure as well as of the characteristic properties of the solitonic cores they harbour. We will first outline their properties individually, then move to the statistical analysis of their properties as a unique population.

\subsubsection{Presence of solitonic cores}

For each zoom-in halo we computed the density profile as described in Sec.~\ref{sec:DP}, finding that a solitonic core is present in the innermost region of all haloes. The soliton density profiles fit well with the approximated analytical function Eq.\ref{eq:soliton}, flattening the typical central density divergence of $\Lambda$CDM haloes. 

\begin{figure}
\includegraphics[width=\columnwidth,trim={0.4cm 0.4cm 0.2cm 0.3cm},clip] {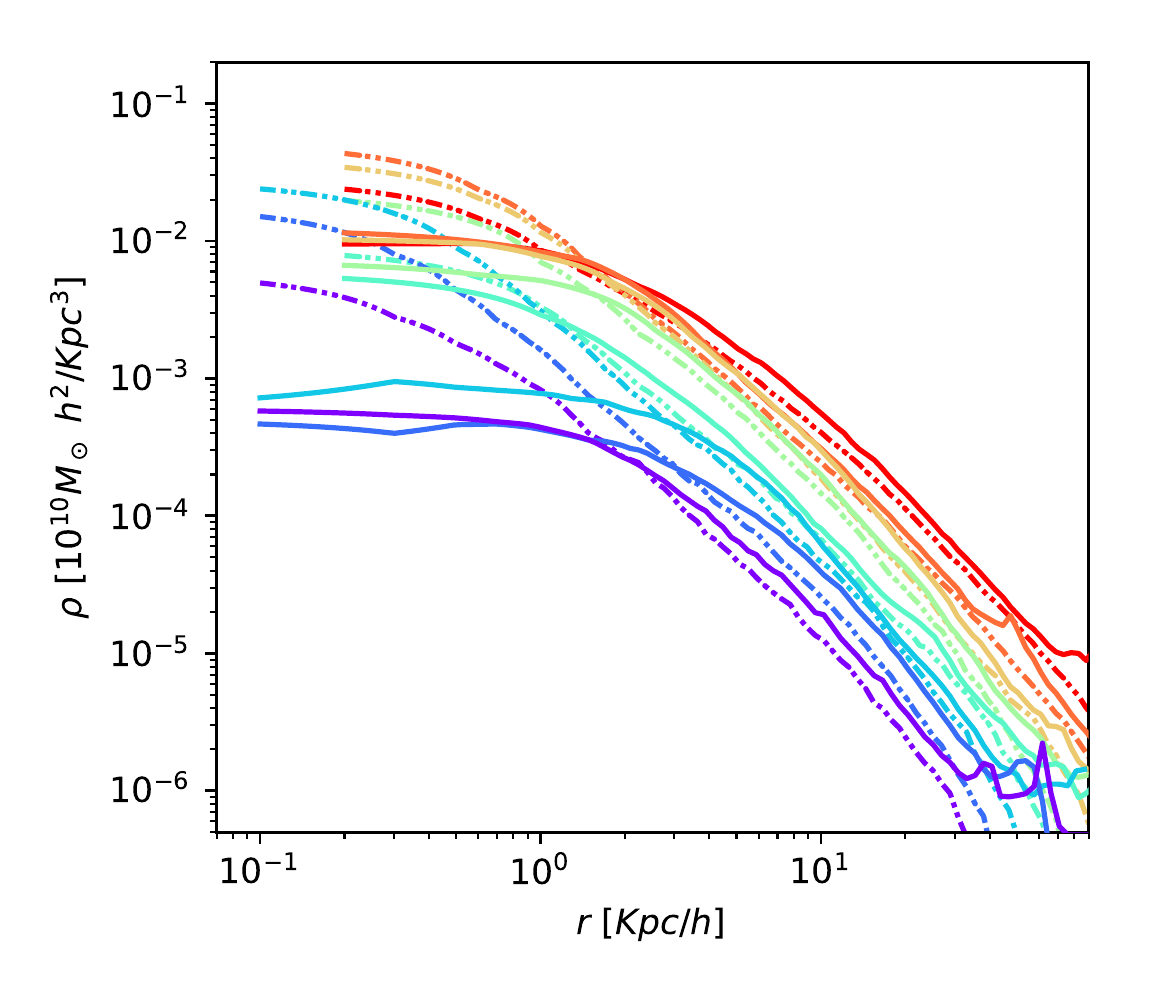}
\includegraphics[width=\columnwidth,trim={0.3cm 0.5cm 0.4cm 0.5cm},clip] {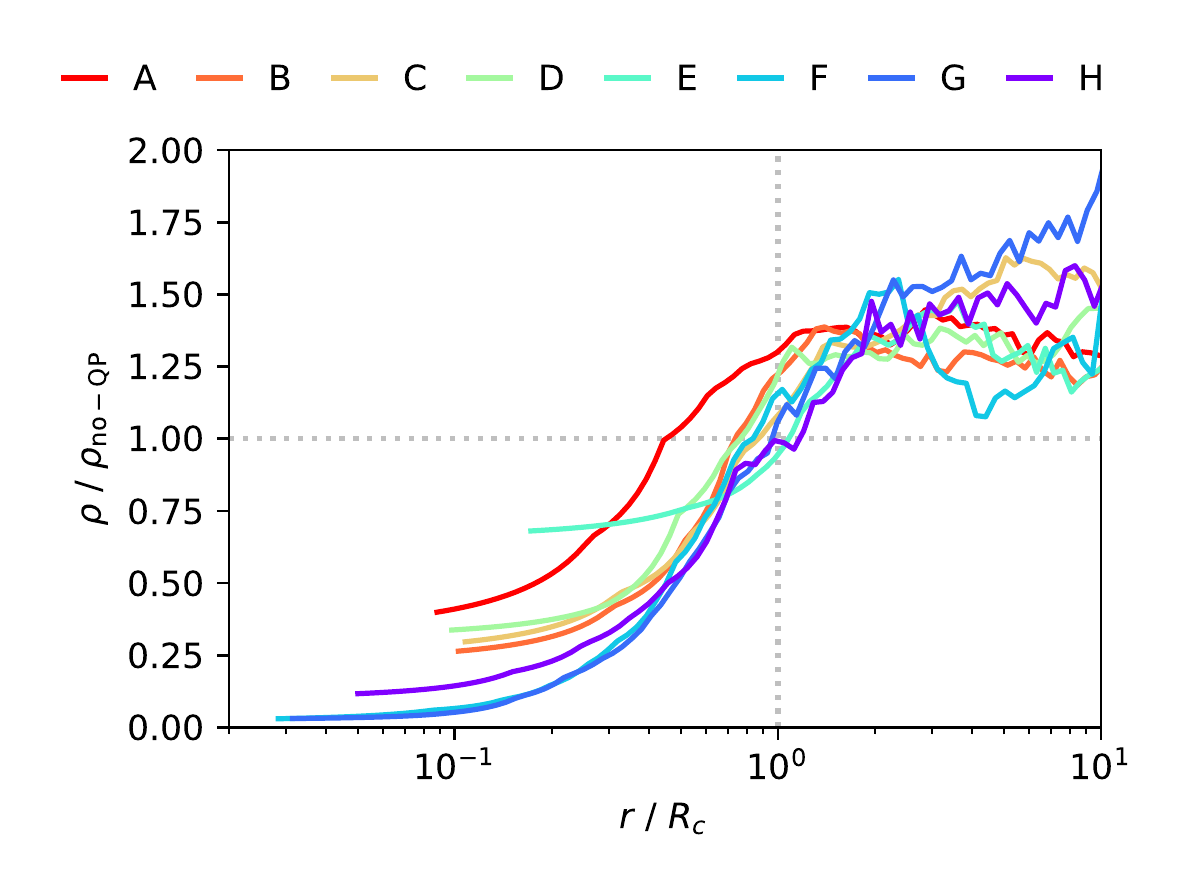}
\caption{\textit{Upper panel}: density profiles of zoom-in haloes at $z=0$ (solid lines) and the profiles extracted from the simulation without the QP (dot-dashed lines). \textit{Lower panel}: ratio of the density profiles with and without the QP. In order to compare systems of different sizes, the radial distance is rescaled by the respective core radii $R_c$.}
\label{fig:profile_noqp}
\end{figure}

In the top panel of Fig.~\ref{fig:profile_noqp}, the density profiles of the zoom-in haloes simulated at $z=0$ with the full FDM dynamics (solid lines) are displayed along with the ones obtained from a set of identical simulations run by switching off the the QP contribution (dot-dashed lines). By comparing the density profiles system-by-system, it is possible to unequivocally attribute the suppression in the central region --~and eventually, the core formation~-- to the QP. This first result establishes that an adequate treatment of the QP is necessary and the suppression of the matter power-spectrum in the initial conditions alone is not enough to correctly reproduce the evolution of FDM systems down to the core level. Although specific to \AG, this results also confirms that the N-body approach is effective in the representation of FDM collapsed objects.

To visually clarify the effect of the QP on FDM haloes profiles, we gathered the ratios between profiles with and without QP in the bottom panel of Fig.~\ref{fig:profile_noqp}, where the distance from the halo centre is expressed in terms of $\Rc$ to allow for a direct comparison between systems with different size. As expected, the QP efficiently counteracts gravity by pushing mass out of the central region --~process which is more efficient in the less massive haloes than in the more massive ones~-- and the equivalence between the profiles with and without QP is statistically found at a distance $r \sim \Rc$ from the centre.

\begin{figure*}
\includegraphics[width=\textwidth,
trim={0.5cm 0.5cm 0.5cm 0.5cm}, clip] {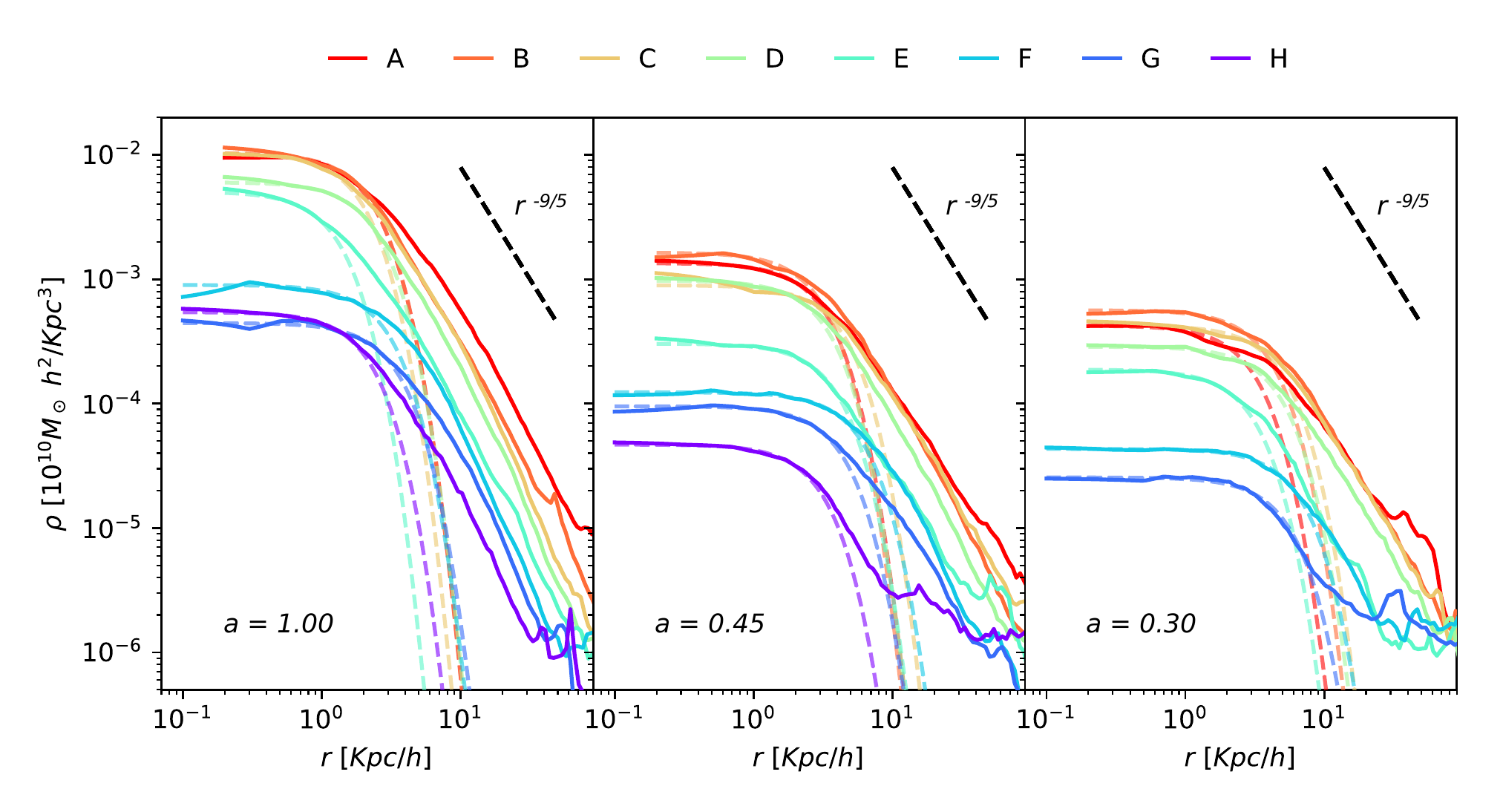}
\caption{Density radial profiles of zoom-in haloes at different redshifts. Dashed lines represent the fitted solitonic core profiles.}
\label{fig:profile_cores}
\end{figure*}

In Fig.~\ref{fig:profile_cores}, density profiles (solid lines) and the solitonic profiles (dashed lines) are displayed at three different redshifts. As it can be seen, the soliton profiles are very well recovered, both for the smallest and newly collapsed haloes --~for which the soliton makes up for a large fraction of the total mass, as observed in the single object collapse test~--, as for haloes at lower redshifts, where the solitonic cores are found to be embedded in a standard NFW-like dark matter halo --~as expected~-- with a smoothly decreasing density profile scaling as $\rho(r) \propto r^{-[1 \div 3]}$. In particular, we find that the scaling $r^{-9/5}$ \citep[observed also by][ and depicted in the figure as a black dashed line]{Eggemeier19} fits particularly well the outer part of the density profile.

\begin{figure*}
\includegraphics[width=0.246\textwidth,trim={0cm 0cm 0cm 0cm},clip]{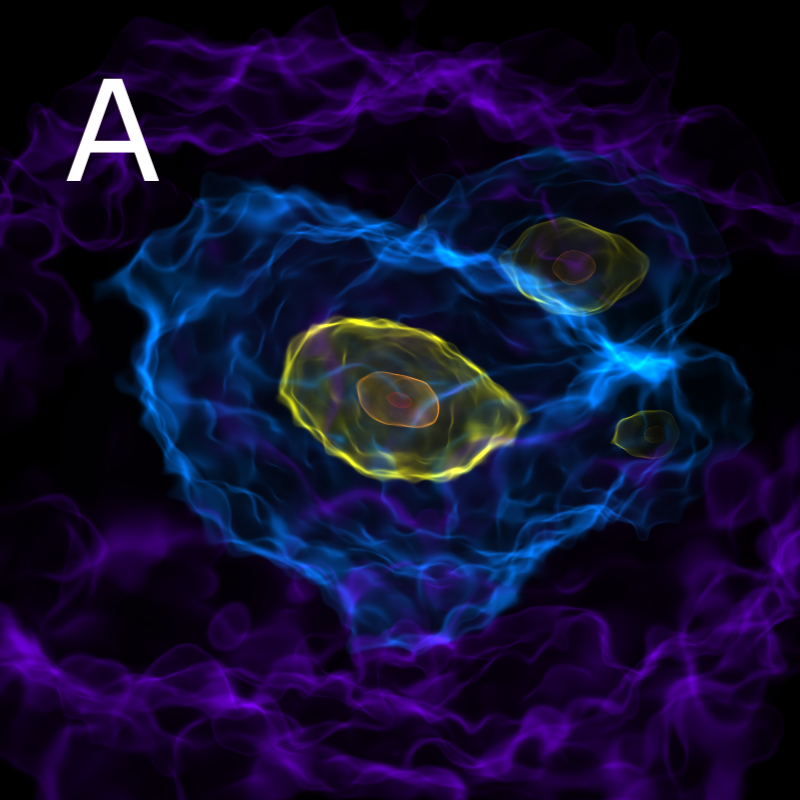}
\includegraphics[width=0.246\textwidth,trim={0cm 0cm 0cm 0cm},clip]{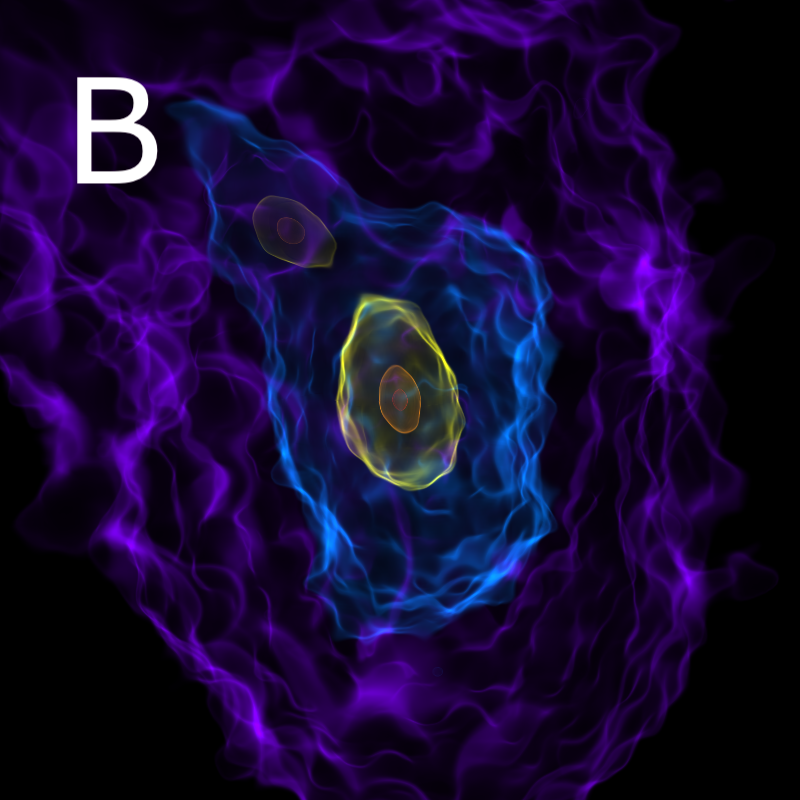}
\includegraphics[width=0.246\textwidth,trim={0cm 0cm 0cm 0cm},clip]{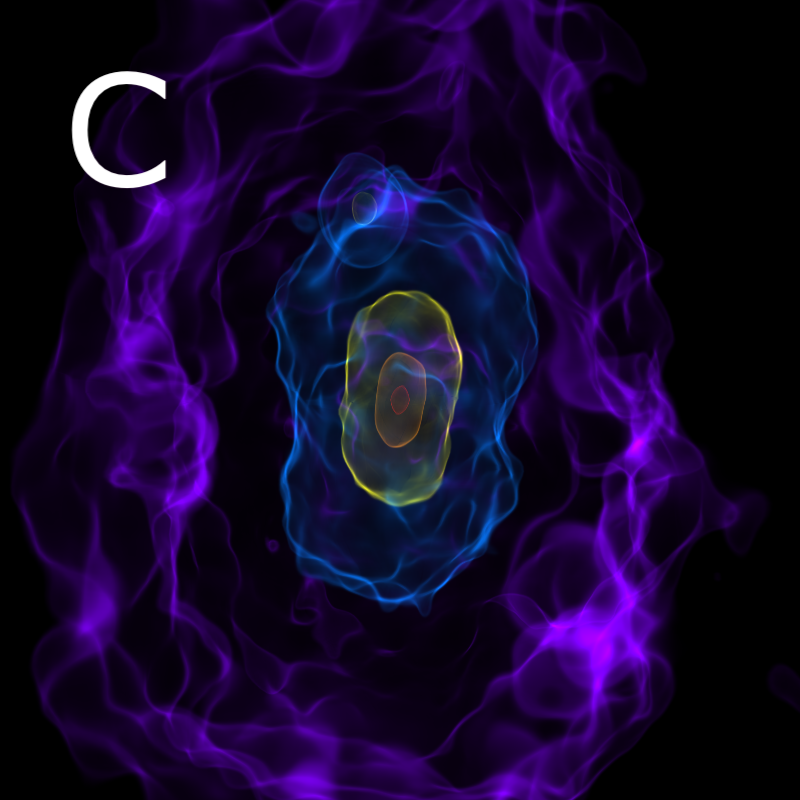}
\includegraphics[width=0.246\textwidth,trim={0cm 0cm 0cm 0cm},clip]{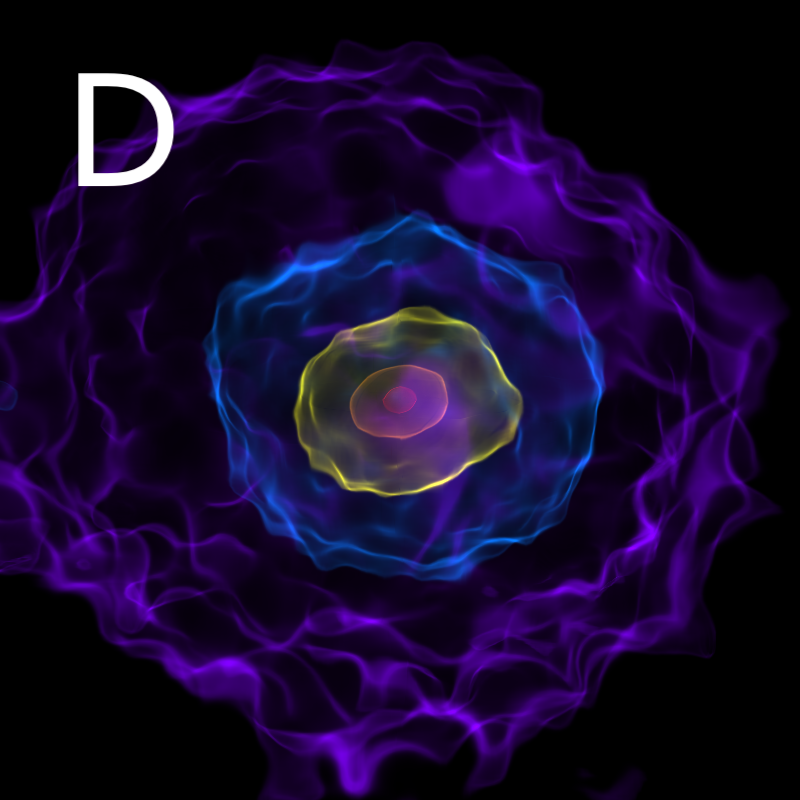} \\
\includegraphics[width=0.246\textwidth,trim={0cm 0cm 0cm 0cm},clip]{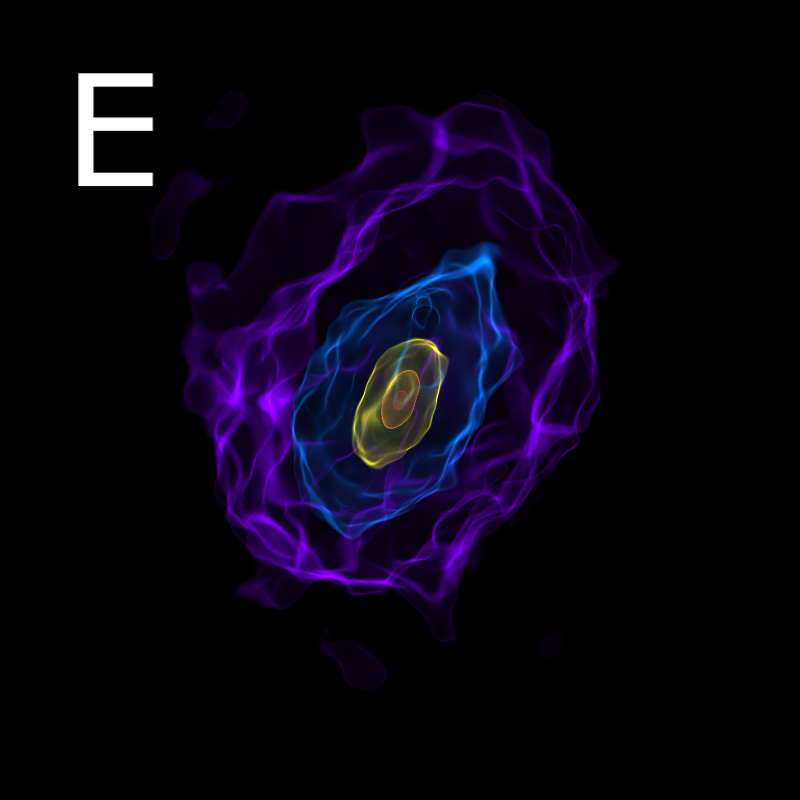}
\includegraphics[width=0.246\textwidth,trim={0cm 0cm 0cm 0cm},clip]{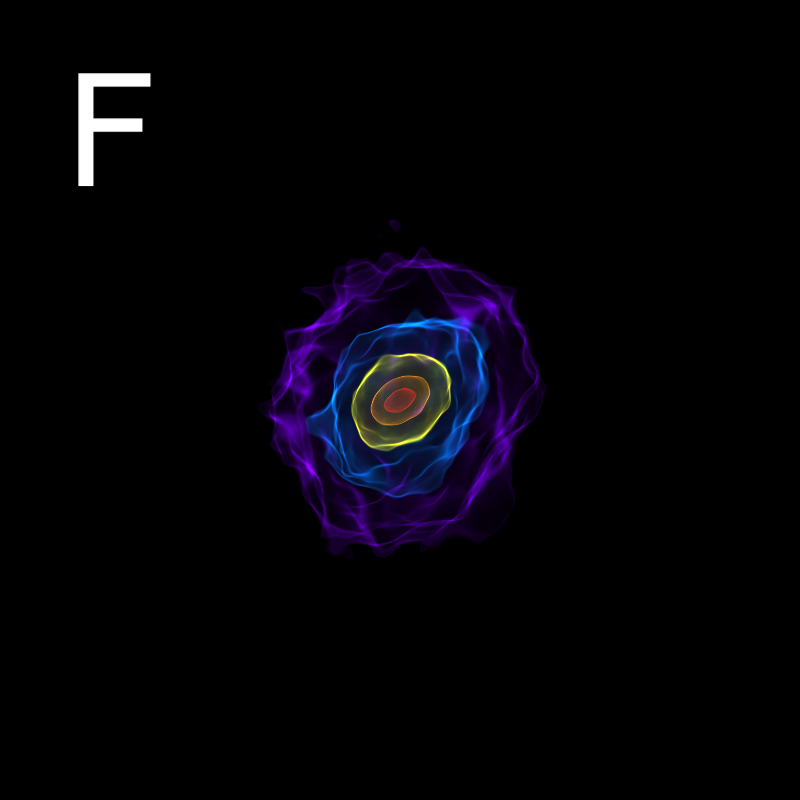}
\includegraphics[width=0.246\textwidth,trim={0cm 0cm 0cm 0cm},clip]{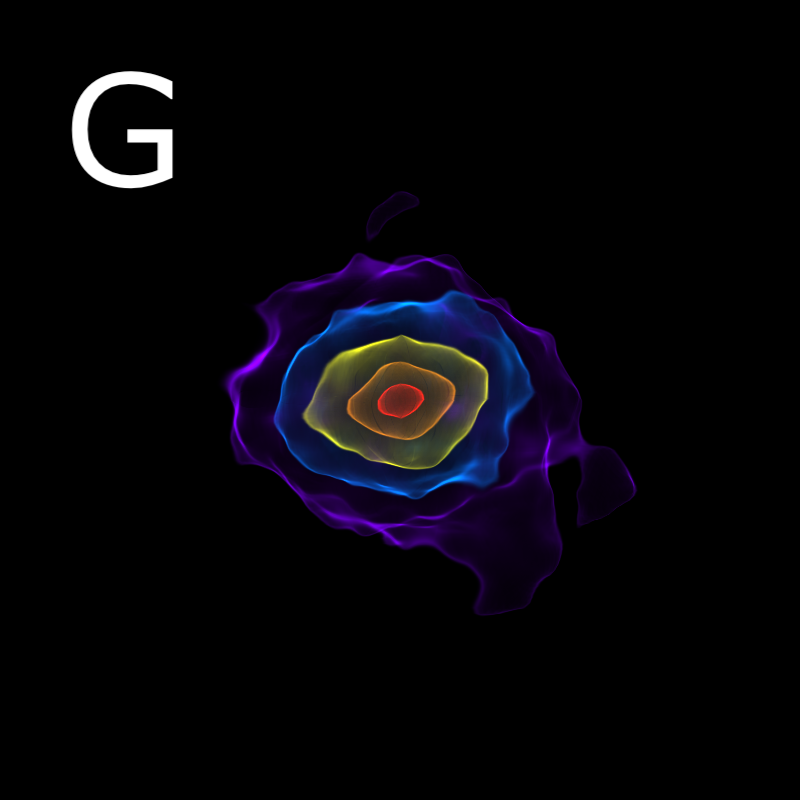}
\includegraphics[width=0.246\textwidth,trim={0cm 0cm 0cm 0cm},clip]{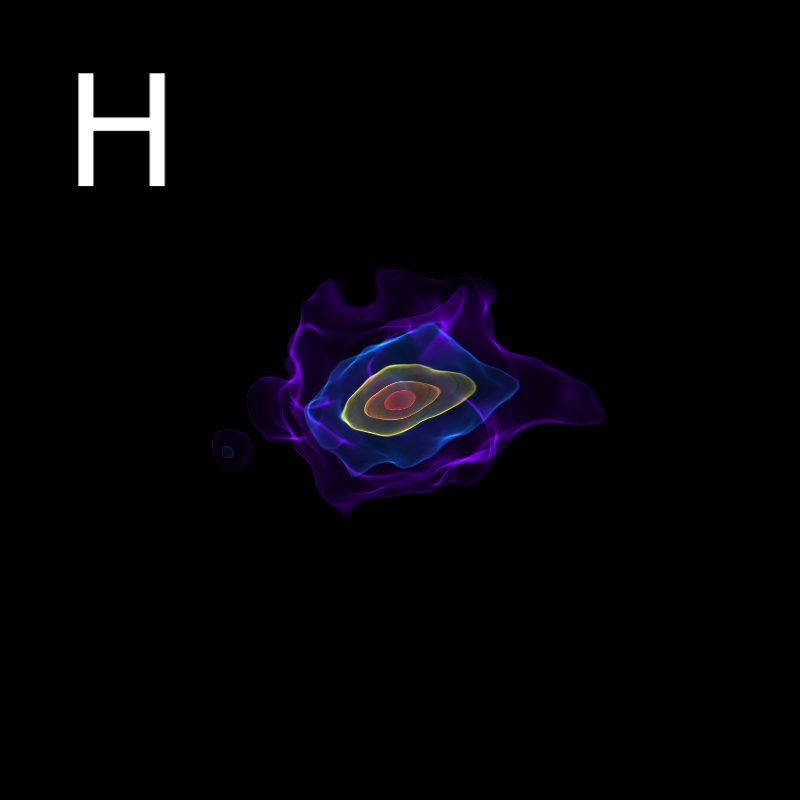}
\caption{3D renderings of the dark matter density of the zoom-in haloes at $z=0$ in a cubic volume of $100 \dimR$ side. The colour scheme identifying the density levels is such that red corresponds to $\rhoc /2$ --~thus ideally representing the core, as defined by Eq.~\ref{eq:Rc}~-- and purple to $\rhovir$; the other colours (orange, yellow and blue) correspond to values between these two, equispaced in logarithmic scale.}
\label{fig:zoom_cores}
\end{figure*}

The properties of the eight zoom-in haloes described in Sec.~\ref{sec:properties} are summarised in Tab.~\ref{tab:ZOOM}, while the 3D rendering of the dark matter density of the eight systems are portrayed in Fig.\ref{fig:zoom_cores} \citep[plotted with the {\small YT} toolkit, see][]{YT}, as extracted from a cubic volume of $100\, \, \dimR$ per side at $z=0$. The coloured manifolds represent the iso-density loci as obtained by mapping the SPH density of particles onto a 3D grid. The colour scheme used to identify the iso-density levels is such that red corresponds to $\rhoc/2$ --~thus ideally representing the core, as defined by Eq.~\ref{eq:Rc}~-- and purple corresponds to $\rhovir$; the other contours (orange, yellow and blue) correspond to densities in between the two, equispaced in logarithmic scale. In this picture, it is possible to appreciate that the least massive and smallest systems are isolated objects in which the core is particularly evident, while the most massive and largest ones host one or few substructures and are characterised by a smaller core relatively to the size of the whole system --~to the point of being barely visible in the picture~--, as quantitatively detailed in Tab~\ref{tab:ZOOM}.

\begin{table*}
\caption{Summary of zoom-in halo and core properties at redshift $z=0$.}
\label{tab:ZOOM}
\center
\begin{tabular}{ccccccccccc}
\hline
\hline
\multirow{2}{*}{Name} & \multirow{2}{*}{$N_\text{part}$} & $M_\text{tot}$ & $\Mvir$ & $\Rvir$ & $\Mc$ & $\Rc$ & \multirow{2}{*}{$s=\dfrac a c$} & $d_\text{off}$ & \multirow{2}{*}{$M_\text{merg}$} & \multirow{2}{*}{$z_\text{form}$} \\
 & & $[10^{10} \dimM]$ & $[10^{10} \dimM]$ & $[\dimR]$ & $[10^{10} \dimM]$ & $[\dimR]$ & & $[\dimR]$ & & \\
\hline
A   & $2383360$    & $65.905$    & $61.705$ & $172.471$ & $0.212$ & $2.348$ & $0.58$ & $6.467$ &  $4.8\%$ & $4$ \\
B   & $1049930$    & $29.033$    & $27.803$ & $132.222$ & $0.217$ & $2.131$ & $0.59$ & $6.785$ & $11.0\%$ & $5$ \\
C   & $505523$     & $13.979$    & $13.667$ & $104.352$ & $0.131$ & $1.966$ & $0.68$ & $1.609$ &  $0.0\%$ & $5$\\
D   & $286746$     & $7.929$     &  $7.795$ &  $86.538$ & $0.156$ & $2.236$ & $0.76$ & $2.526$ &  $0.0\%$ & $5$\\
E   & $147959$     & $4.091$     &  $3.881$ &  $68.588$ & $0.020$ & $1.477$ & $0.55$ & $0.965$ &  $2.3\%$ & $3$ \\
F   & $611880$     & $2.115$     &  $2.050$ &  $55.440$ & $0.044$ & $3.198$ & $0.70$ & $0.918$ &  $3.1\%$ & $3$\\
G   & $257709$     & $0.891$     &  $0.832$ &  $41.049$ & $0.034$ & $3.307$ & $0.44$ & $0.662$ &  $0.0\%$ & $2.33$\\
H   & $123644$     & $0.428$     &  $0.389$ &  $31.860$ & $0.010$ & $2.104$ & $0.64$ & $0.719$ & $0.5\%$ & $1.2$\\
\hline
\hline
\end{tabular}
\end{table*}

\bigskip

As described in Sec.~\ref{sec:mergertree}, we selected haloes that formed relatively early with respect to haloes of the same mass range in the \textit{COARSE} simulation, in order to be able to have more data on their evolution. Indeed, the merger tree analysis confirmed that all but two haloes formed at a redshift $z_{\rm form} \geq 3$, with \textit{G} and \textit{H} forming at $z_{\rm form} = 2.33$ and $z_{\rm form} = 1.2$, respectively. It is clear that age correlates positively with mass of a dark matter system, since the oldest haloes have been continuously accreting mass for longer time. Moreover, old systems had a higher chance to take part in a merger event with respect to younger ones, although our selection criteria tend to exclude highly interacting systems by construction.

All haloes are consistent with a solitonic profile from the time of formation onward, with the special exception worth mentioning represented by halo \textit{H}, which forms from the radial collapse of a filament \citep[see e.g.][for an example of FDM non-spherical core solutions]{Mocz19,Mocz19companion,Bar19}. Indeed, its density profile at the redshift of formation $z=1.8$ shows the onset of a solitonic cylindrical ``core'' that then collapses longitudinally, transitioning to a more spherical system by $z=1.2$: for this reason, we will exclude halo \textit{H} data prior to $z=1.2$ from our analysis. For a detailed description of this interesting metamorphosis, we refer the reader to App.~\ref{sec:filament}.

\subsubsection{Statistical analysis: an {\normalfont{agnostic}} approach}

As described in Sec.~\ref{sec:fdm_sr}, the core properties should \textit{statistically} satisfy SRI and SRII if the related assumptions are verified. 

\begin{figure*}
\includegraphics[width=1.0\textwidth,trim={0.5cm 0.5cm 0.6cm 0.5cm},clip] {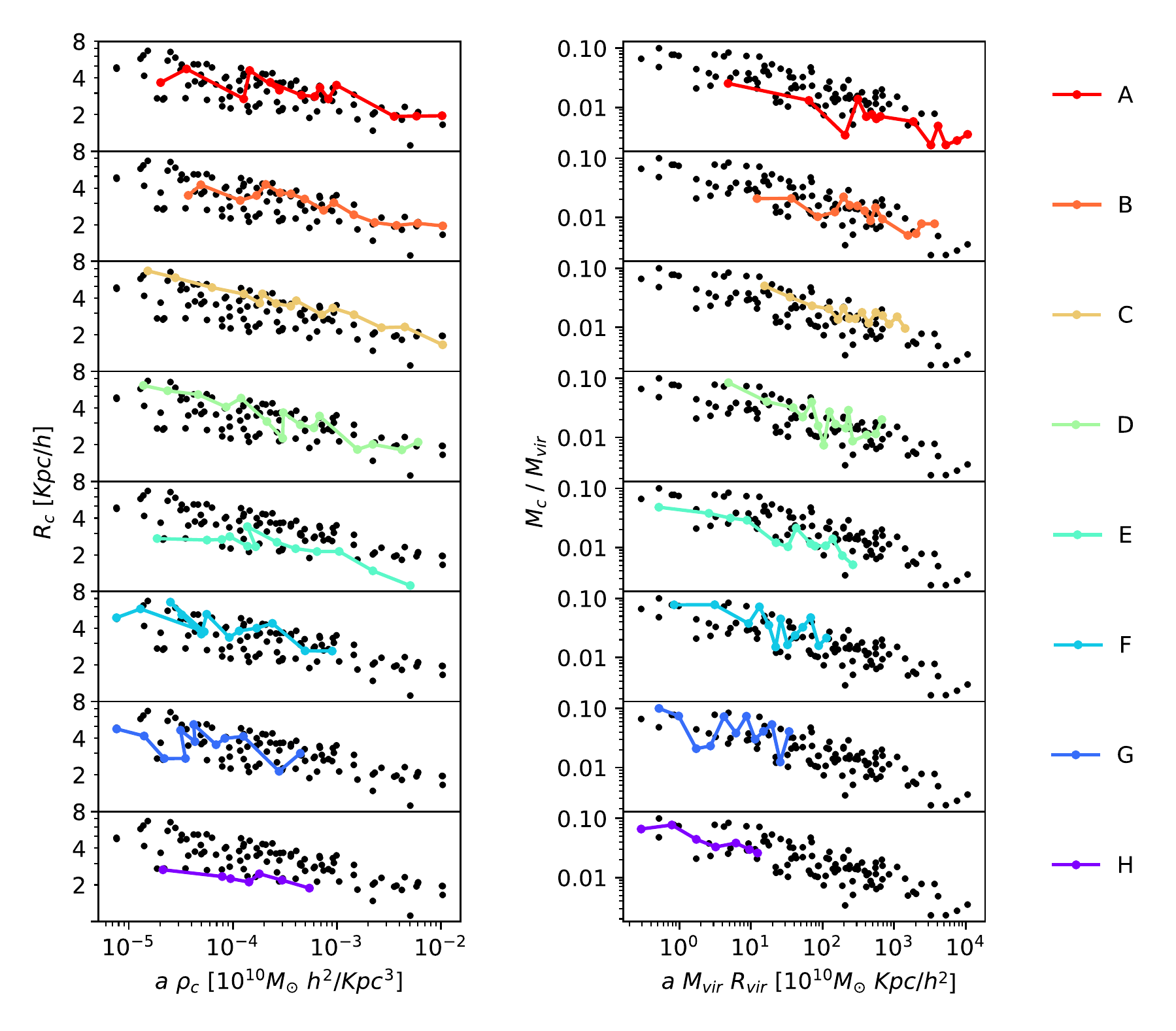}
\caption{Properties of zoom-in haloes, gathered in $(\rhoc,\Rc)$ (left panels) and $(\Mc/\Mvir,\Mvir\Rvir)$ (right panels) parameter spaces, where black data points represent all haloes at all redshifts. The datapoints related to a particular zoom-in halo are highlighted row-wise as coloured lines. In both parameter spaces, time flows from the top left to bottom right; however, the exact time evolution --~as described by Eq.~\ref{eq:SR}~-- depends on the scaling exponents.}
\label{fig:halowise}
\end{figure*}

To place the properties of each zoom-in halo within the context of the whole population, in Fig.~\ref{fig:halowise} we collect a series of scatter plots gathering the observables $(\Rc, \rhoc)$ (\textit{left} column) and $(\Mc/\Mvir, 1/\Mvir\Rvir)$ (\textit{right} column) of all zoom-in haloes at all redshifts. Black points represent independent measurements of all haloes at different redshifts and coloured curves visually highlight the evolution of each single halo in redshift row-wise.

Gathered in this fashion, the distribution in these two property spaces is qualitatively consistent with a power-law, both \textit{individually} as single  systems evolving in time and \textit{collectively} as a whole population. Moreover, it is possible to see that the most massive systems statistically harbour the smallest cores --~with respect to the total population~-- in terms of absolute size as well as in mass, relatively to the virial mass.

To quantify these features, we perform a statistical analysis based on power-law fitting and bootstrap re-sampling. The total number of data points $N_{\rm sample}$ in our sample is $104$ --~slightly less than $8$ haloes times $15$ redshifts available, since the smallest haloes were not yet formed at the earliest redshifts~--, so it allows for a safe \textit{bootstrap} procedure to estimate the best values for the scaling relations parameters and their confidence levels. At this stage, we do not differentiate haloes by redshift or any other property, thus implicitly assuming that SRI and SRII are universally valid.

In practice, we performed a logarithmic fitting analysis for each random draw of the bootstrap procedure --~the total number of draws is $N_{\rm sample}^2$, performed by substitution~-- to obtain the parameters $(\kappa,\mu,\tau,\eta)$ of SRI and SRII as from Eq.~\ref{eq:SR}. We then build an occurrence statistics to study the distribution of such parameters and extract confidence regions. We performed two fitting analyses in parallel: one fixing the scaling exponent as in \citet{Schive14} and one allowing the exponents $\mu$ and $\eta$ to vary. In the following discussion, we will use the subscript $\mu$ and $\eta$ for the $\kappa$ and $\tau$ parameters resulting from the varying exponents analysis, while we will specify the values of the subscript when referring to the results obtained by fixing the exponents --~as e.g. $\kappa_{1/4}$ obtained by fixing the exponent $\mu=1/4$~--.

\begin{figure*}
\includegraphics[width=\textwidth,trim={0cm 0.5cm 0cm 0.5cm},clip] {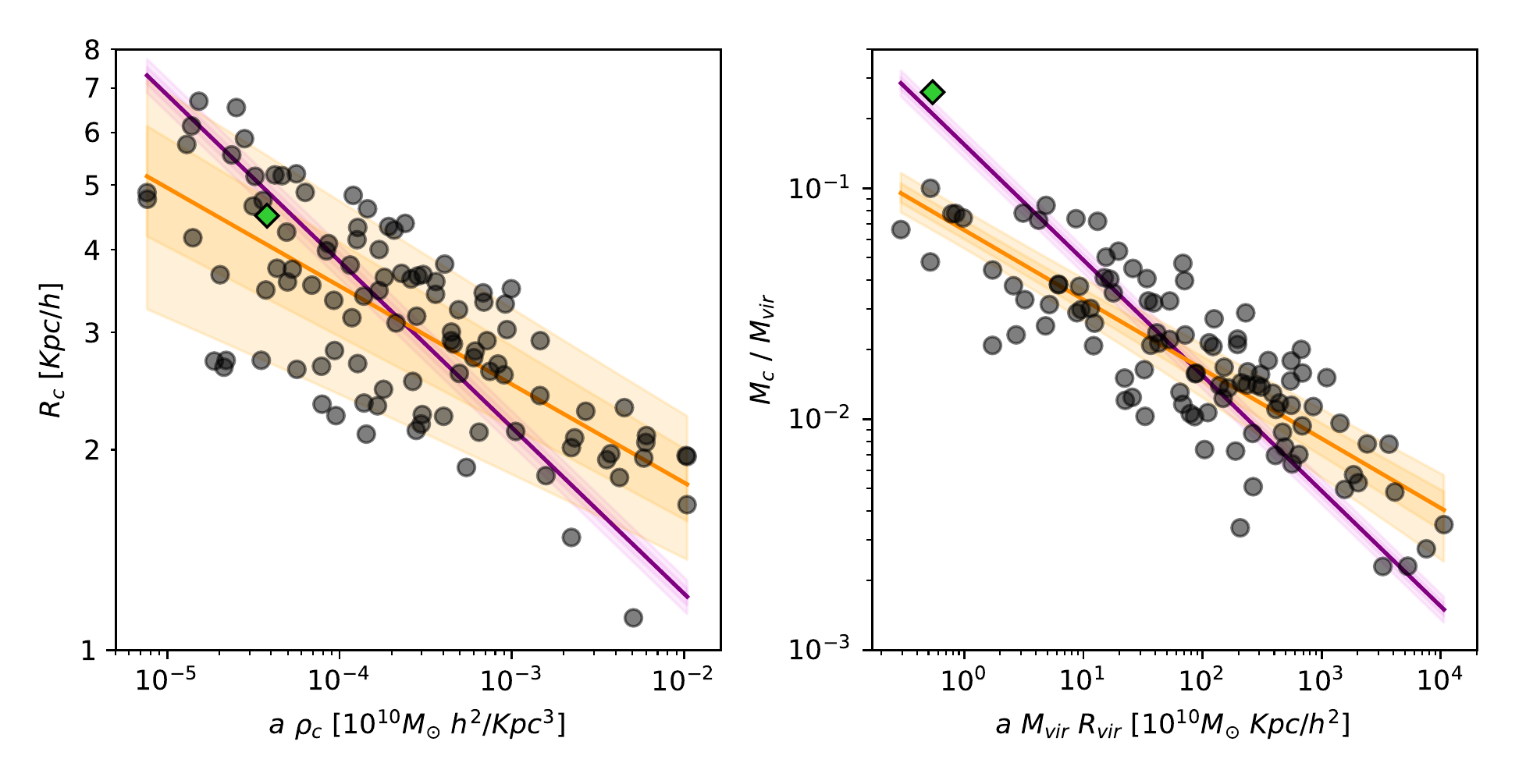}
\includegraphics[width=\textwidth,trim={0.4cm 0.5cm 0.4cm 0.4cm},clip]{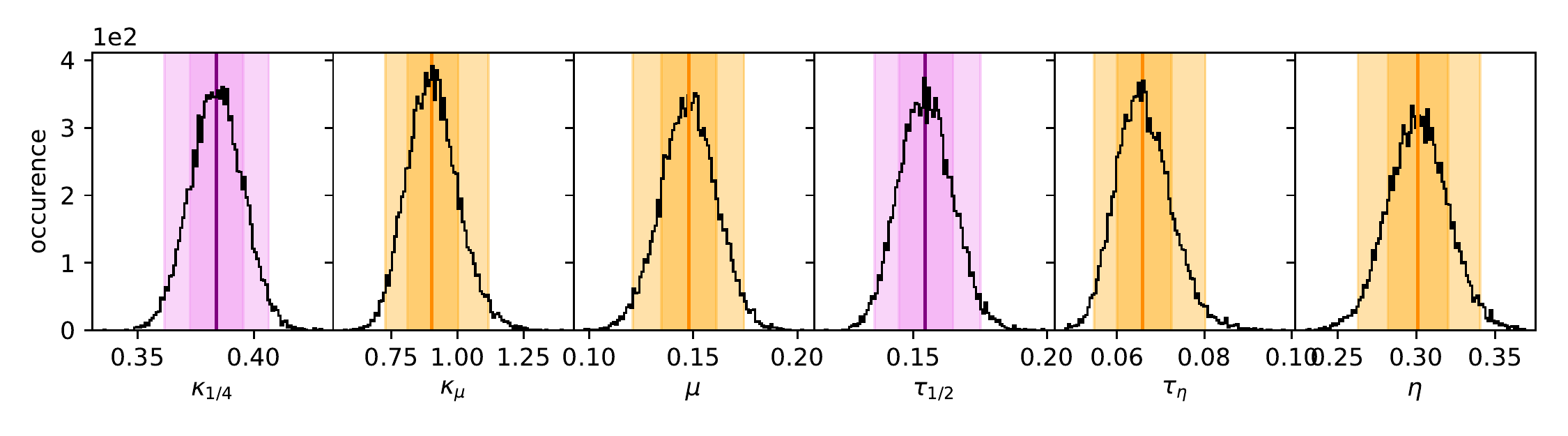}
\caption{Properties of the zoom-in haloes at different redshifts (top panels), portrayed in the parameter spaced used to fit SRI (top left panel) and SRII (top right panel) with the bootstrap statistical approach, with the addition of the data point related to the single object test (green diamond). The resulting parameter distributions are displayed in the \textit{lower} panels). In all panels, solid coloured lines represent the value of best fit and shaded areas represent the $68\%$ and $95\%$ percentile-equivalent confidence region of both the fixed (magenta) and varying (orange) exponent agnostic approach.}
\label{fig:bootstrap}
\end{figure*}

In the two upper panels of Fig.~\ref{fig:bootstrap}, we again display the distribution of all haloes properties as in Fig.~\ref{fig:halowise}, together with the results of the fitting analysis. The best parameter values describing the power-laws obtained with fixed and varying exponents (purple and orange, respectively) are plotted as solid lines and are shown alongside their $68\%$ and $95\%$ confidence regions, depicted as colour-matched shaded areas.

The parameter distributions obtained through the bootstrap procedure are shown in the lower panels of Fig.~\ref{fig:bootstrap} as histograms, where again the solid vertical lines represent the best values --~taken as the median~-- and the shaded areas the $68\%$ and $95\%$ confidence regions. Note that the median value is consistent with the mode and the overall distributions are quite symmetrical in all cases. We summarise the results in Tab.~\ref{tab:summary}, where each parameter value is accompanied by the $68\%$ percentile confidence range. Due to the almost symmetrical nature of these distribution, the $68\%$ confidence ranges are expressed in symmetric fashion for simplicity and approximated by excess; in the following, we will refer to these values as $\sigma$ when comparing different values of the SRI and SRII parameters.

As a methodological note, let us remark that these confidence ranges are not to be confused with the standard deviations associated with the sample: the former is a measure of how probable the best fit values are as compared to another random sample extracted from the true (unknown) population, while the latter is a measure of the intrinsic spread in the dataset with respect to the best values. From a statistical point of view, the relation between these two quantities is analogous to the relation of the standard error of the mean to the standard deviation of a dataset. As an example, the standard deviations in the varying exponents case are $\Delta \Rc \sim 1.4\, \,  \dimR$ and $\Delta \Mc/\Mvir \sim 0.02$: these estimates are a proxy for the statistical variability of the properties of a typical object and the ones predicted by SRI and SRII identified by the best values.

\begin{table*}
\centering
\caption{Values of the parameters of SRI and SRII of Eq.~\ref{eq:SR} obtained through bootstrap sampling with different strategies.}
\label{tab:summary}
\begin{tabular}{ccc|cc|cccc}
\hline
\hline
\multicolumn{2}{c}{Sample Restrictions} & \multirow{2}{*}{$N_{\rm sample}$} & \multirow{2}{*}{$\kappa_{1/4}$} & \multirow{2}{*}{$\tau_{1/2}$} & \multirow{2}{*}{$\kappa_\mu$} & \multirow{2}{*}{$\mu$} & \multirow{2}{*}{$\tau_\nu$} & \multirow{2}{*}{$\eta$} \\
$s$ & $d_{\rm off} / \Rvir$ & & & & & & & \\
\hline
- & - & $104$ &$0.384\pm0.011$ & $0.155\pm0.010$ & $0.907\pm0.095$ & $0.148\pm0.013$ & $0.066\pm0.006$ & $0.301\pm0.019$ \\
$>0.4$ & $<0.10$ & $59$ & $0.424\pm0.013$ & $0.189\pm0.014$ & $0.750\pm0.094$ & $0.176\pm0.017$ & $0.077\pm0.013$ & $0.324\pm0.031$ \\
$>0.6$ & $<0.07$ & $25$ & $0.423\pm0.019$ & $0.195\pm0.023$ & $0.707\pm0.106$ & $0.185\pm0.020$ & $0.086\pm0.028$ & $0.319\pm0.071$ \\
\hline
\hline
\end{tabular}
\end{table*}

Regarding the distribution of the $\mu$ exponents of SRI in this agnostic approach, we find that the value $\mu=1/4$ lies at $\sim 7.5\sigma$ from the most probable value obtained in the case where the exponent $\mu$ is free to vary: the value $\mu=1/4$ thus seems inconsistent with the data, but a lower exponent is preferred. Turning to the exponent $\eta$ of SRII, our best value is $\sim 1.5 \sigma$ away from the value $\eta=1/3$ as found in \citet{Mocz17}, disfavouring the value $\eta=1/2$ obtained by \citet{Schive14} which lies $\gtrsim 10\sigma$ away from the mean.

Furthermore, the results previously obtained in the single object collapse test are plotted as a green diamond in Fig.~\ref{fig:bootstrap}. These data points of non-cosmological origin are interestingly consistent with the scaling relations with fixed exponents, in contrast with what found when exponents are free to vary: this result suggests that scaling relations may be altered by the cosmological context in which haloes and cores form and co-evolve.

Indeed, the inconsistency between our datasets and the theoretical predictions --~especially regarding SRI~-- might seem troublesome at first sight, but let us recall that these results are obtained considering the total sample of observables without taking into account dynamical and morphological information of the host haloes, that we are going to include in the analysis in the following Sections.

\subsubsection{Dynamical and morphological information of the host haloes}

\begin{figure*}
\includegraphics[width=\textwidth, trim={0cm 1.2cm -0.32cm 0.4cm}, clip] {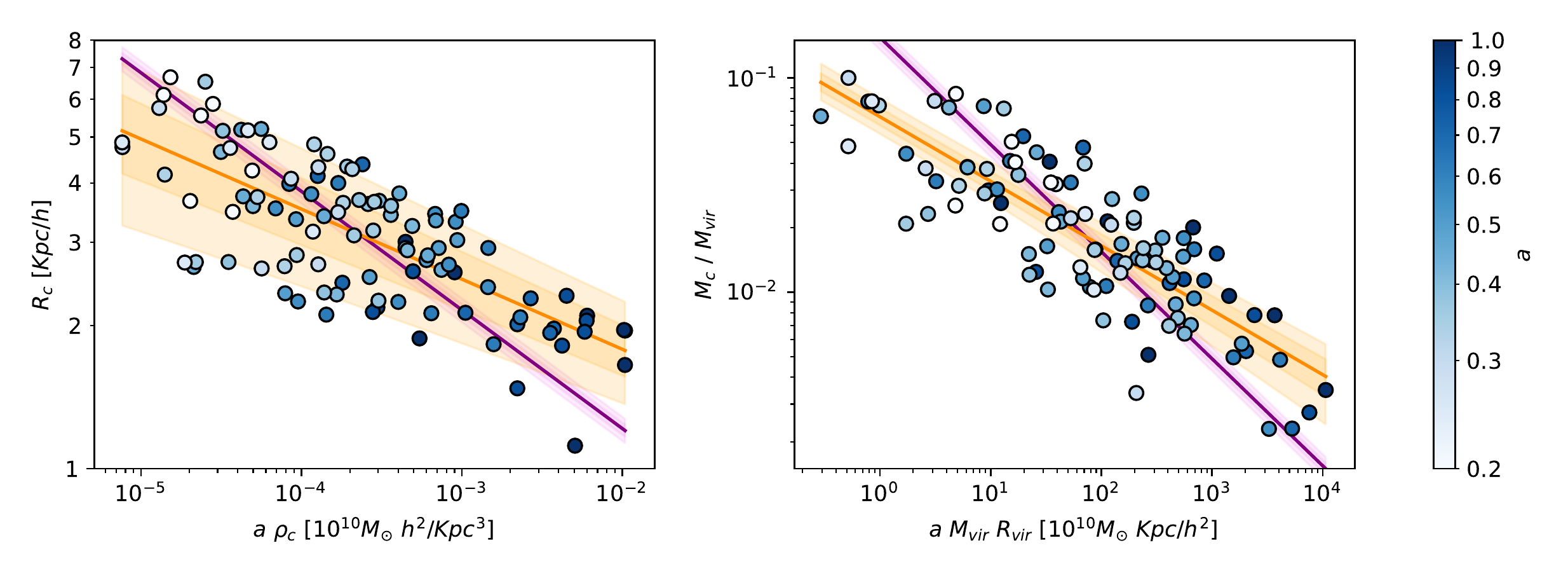}
\includegraphics[width=\textwidth, trim={0cm 1.2cm 0.13cm 0.4cm}, clip] {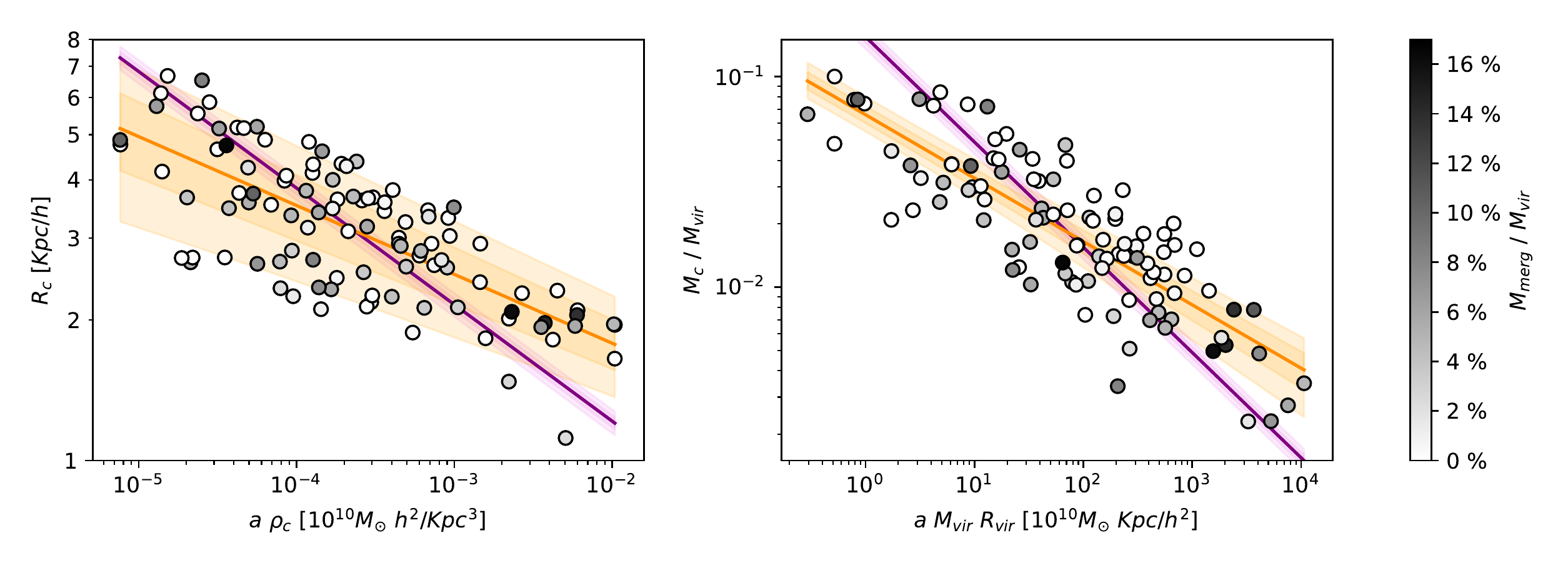}
\includegraphics[width=\textwidth, trim={0cm 1.2cm -0.32cm 0.4cm}, clip] {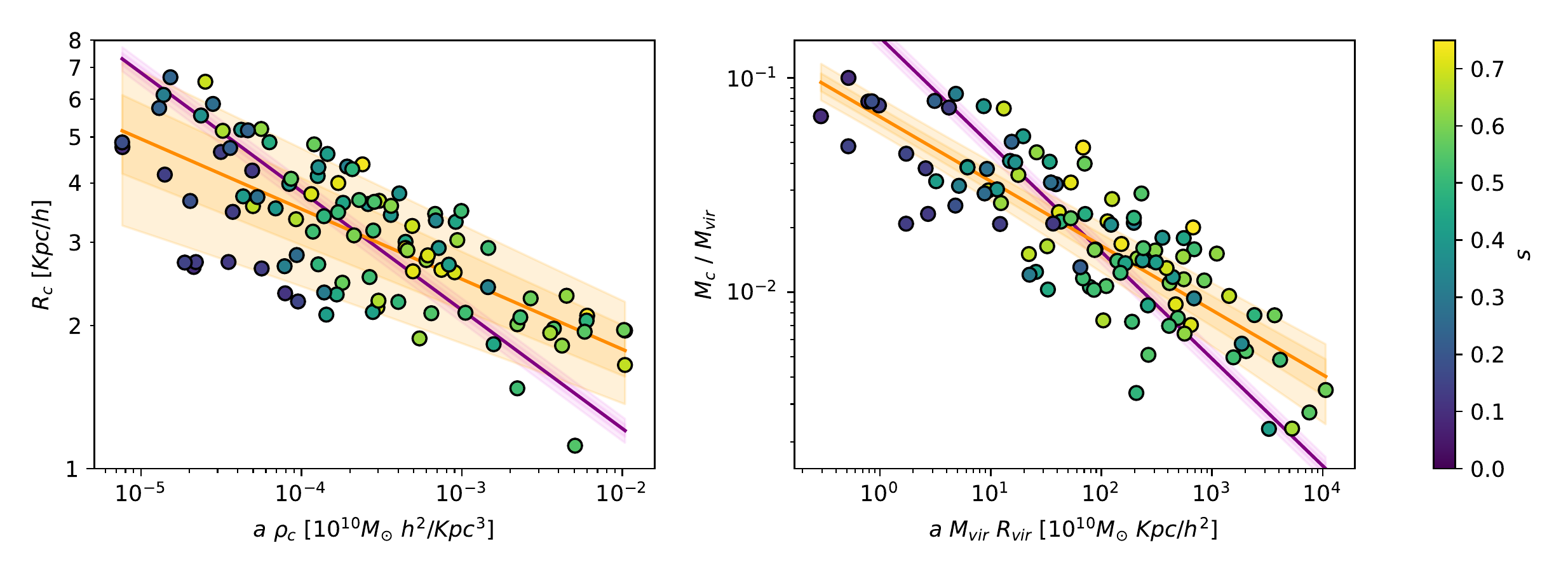}
\includegraphics[width=\textwidth, trim={0cm 0.5cm 0cm 0.4cm}, clip] {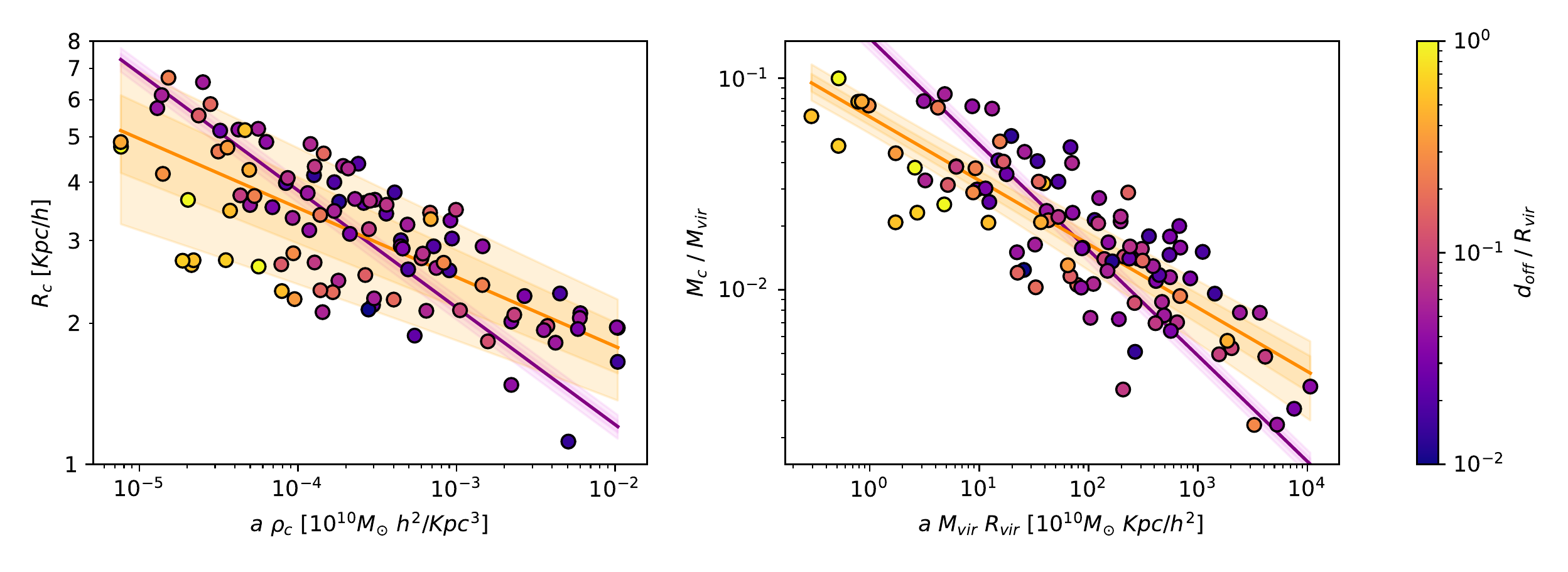}
\caption{Replicas of the properties of the zoom-in haloes at different redshifts, as in Fig.~\ref{fig:bootstrap}. Each data point is here coloured according to an additional property. From top to bottom: scale factor $a$, mass gained via merger $M_{\rm merg}$, sphericity $s$ and centre offset $d_{\rm off}/\Rvir$.}
\label{fig:observables}
\end{figure*}

As we discussed in Sec~\ref{sec:fdm_sr}, SRI and SRII both rely on the assumption of spherical symmetry and relaxed dynamical state of the system at all times. In this Section, we check the validity of these assumptions for our dataset by investigating the observables discussed in Sec.~\ref{sec:properties}. In Fig.~\ref{fig:observables}, we replicate the results in the same parameter space as in Fig.~\ref{fig:bootstrap}, with data points in each row are here colour coded according to a different property: from top to bottom, the colours are representative of the scale factor $a$, the mass accreted via merger $M_{\rm merg}$, the sphericity $s$ and the centre offset (in units of the halo virial radius) $d_{\rm off}/\Rvir$.

\bigskip

Starting from the first row depicting the scale factor distribution of core and halo properties, we expect a general trend to be evident: as a system evolves, the density of the core increases and its radius gradually shrinks, so that the core mass decreases with respect to the virial mass. In fact, as SRI and SRII are explicit functions of $a$, evolution in time can be identified in both the spaces with a flow of data points from the upper left corner toward the bottom right corner, with the direction of the flow depending on the specific exponent assumed to be true. In addition, the global time evolution arises also as an implicit (and non-linear) dependence of each halo property --~e.g. the implicit time dependence $\rhoc(a)$ and $\Mvir(a) \Rvir(a)$ in our case~-- which complicates the overall description of the relation between scale factor and the core/halo properties. 

For SRI, it seems that there is indeed a statistical correspondence between the scale factor and the locus where different systems are found along the relation. Specifically, different systems at early times have rather similar values of $(\rhoc,\Rc)$ and coherently evolve in time, though with increasing spread as time progresses. For SRII, instead, a global scale factor correlation with $(\Mc/\Mvir,\Mvir \Rvir)$ is not evident, as different systems at the same redshift occupy distant positions along the power-law curve with no clear trend; however, comparing this information with Fig.~\ref{fig:halowise}, it is possible to see that the time evolution of all systems is \textit{individually} consistent with a progression along the power-law curve --~with the $\Mc / \Mvir$ ratio becoming smaller and smaller in time~--, yet the starting and ending points of the trajectories of more massive systems are systematically shifted along the power-law towards smaller $\Mc / \Mvir$ ratios, so that a \textit{global} correspondence between the scale factor and the distribution of datapoints in the SRII parameter space it is not particularly evident.

\bigskip

Looking at the panels in the second row, we note that the merger analysis shows that the large majority of systems have none or negligible mass contribution accreted via merger with respect to their virial mass. In fact, for $43$ out of $104$ total datapoints we have that $M_{\rm merg}$ is identically zero while the overall mean and maximum values of $M_{\rm merg} / \Mvir$ are $\sim 2.8\%$ and $\sim 16.5\%$, respectively, with
only $7$ datapoints having a contribution $M_{\rm merg} / \Mvir > 10\%$.
The mass contribution has no clear correlation with the SRI or SRII distribution. Nevertheless, it is possible to note that the systems with a highest mass contribution seem to be found in two main regions in the parameter spaces: one consistent with the young and small systems in the upper left corner of the plots --~for which a single merger makes up for a larger contribution, due to the small virial mass~-- and for old and big systems in the opposite corner --~for which, instead, there is a higher incoming mass contribution, as resulting from a larger number of major merger events~--. In the end, no conclusive and univocal link can be drawn between mass accreted via merger and any scaling relation in our dataset; as detailed in Sec.~\ref{sec:mergertree}, this comes as no surprise, as our selection procedure for the zoom-in haloes introduced a bias in our sample, effectively excluding strongly interacting systems. We will overcome this limitation and thoroughly investigate the role of mergers in the onset of scaling relations in an upcoming companion paper \PARTII.

\bigskip

Moving on to the halo shape in the third row of Fig.~\ref{fig:observables}, we find that the distribution of halo sphericity is extremely interesting, as it clearly correlates with the parameter space distribution of both SRI and SRII: the least spherical objects (i.e. the darkest data points in the figure) mostly placed in the left side of the plot are the ones that deviate the most with respect to the SRI with $\mu=1/4$ found by \citet{Schive14}, whose core radius and density (and mass) appear to be underestimated, probably as a consequence of the invalid assumption of sphericity. It is visually clear that these systems are responsible for a systematic deviation of the bootstrap results from SRI in the \textit{agnostic} case. With respect to SRII, the systems with low sphericity occupy a specific region in the parameter space --~in particular, the upper left corner characterising the youngest haloes with the highest $\Mc / \Mvir$ ratio, consistently with the shape evolution of standard structure formation \citep[][]{Zeldovich_1970}~-- and, also in this case, they show a systematic deviation from the $\eta=1/2$ power-law trend of \citet{Schive14}.

\bigskip

The least spherical systems are also the ones that exhibit a large centre offset, shown in the fouth and last row. In fact, the centre offset distribution seemingly overlaps with the sphericity one, as they both are pivots for the dynamical state of the halo\footnote{The tight correlation between sphericity and the logarithm of the centre offset can be indeed measured, resulting in a correlation factor of $\sim-0.75$.}. The systems that are least spherical and have the largest centre offset represent a sub-population that occupies a specific region in the left side of the two property spaces far from the power-laws with exponents found by \citet{Schive14}. At the same time, they are the smallest simulated haloes at the highest redshifts available, for which we expect a dynamical state far from relaxed. Yet, these objects are real physical structures --~just in an early stage of their evolution~-- and might host visible galaxies that would be indeed present (and observable) in a FDM-dominated Universe, so that they should not be excluded from the assessment of general scaling relations.

\subsubsection{Statistical analysis: a {\normalfont{biased}} approach}

The core and halo properties we detailed above give us an important insight on the dynamical state and morphological features of each system. As we discussed in Sec.~\ref{sec:fdm_sr}, this information plays a significant role as a check on the assumptions supporting SRI and SRII. In particular, we noted that the sphericity $s$ and the the centre offset in units of the virial radius $d_{\rm off} / \Rvir$ of systems seem to correlate with the distance of system properties from the ones predicted with the best fit obtained with an agnostic approach. Moreover, these two quantities are directly linked to the spherical symmetry and relaxed dynamics assumptions, so including them in the fit analysis is of great significance.

To quantitatively describe the connection between the sphericity $s$, the centre offset in units of the virial radius $d_{\rm off} / \Rvir$ and the scaling relations, we repeated our bootstrap analysis on two subsamples of haloes, defined by an increasingly stringent cutoffs on sphericity and centre offset. The first subsample is characterised by $s>0.4$ and $d_{\rm off} / \Rvir < 0.10$ and the second one by $s>0.6$ and $d_{\rm off} / \Rvir < 0.07$. The values for the sphericity cut were chosen to exclude the lower end of the sphericity distribution while keeping $N_{\rm sample}$ high enough to be statistically relevant, taking into account that the maximum sphericity in the sample is $\lesssim 0.8$. The same is valid for the centre offset cutoff values, that we chose also consistently with the value ($0.07$) used in \citet{Neto_etal_2007} to identify relaxed objects.

\bigskip

The parameters obtained by the bootstrap analysis, summarised in Tab.~\ref{tab:summary}, show how the exponent $\mu$ systematically (and significantly) shifts towards the theoretically predicted value $\mu=1/4$ for spherically symmetric systems, with best-fit values of $\mu=0.176$ and $\mu=0.185$ for the two increasingly selective subsamples --~i.e. with a $\sim 30\%$ increase with respect to the \textit{agnostic} case~--. Having in mind that our sample does not include haloes with sphericity higher than $s>0.8$ --~which are very rare among FDM structures \citep[as also seen in][]{Nori19}~--, this result suggests that the exponent $\mu=1/4$ of SRI may only represent an asymptotic limit that can be retrieved from systems for which the spherical symmetry assumption is approximately valid. The perfect sphericity of all dark matter haloes, as a matter of fact, is never realised --~not even at low redshift~-- in a realistic cosmological context, where the sphericity distribution is continuous and broad, encompassing a large number of less spherical systems. From a different point of view, we can say that it is possible --~and reasonable~-- to interpret the inconsistency between the youngest and least spherical system with the exponent $\mu=1/4$ of SRI as a sign of the excited state of these cores, as they may have not yet fully stabilised in the solitonic ground-state \citep[see e.g. the excited state of a FDM core observed in][]{Veltmaat18}.

The exponent $\eta$ related to SRII appears to be less affected by these sample restrictions, shifting the exponent towards $\sim15\%$ higher values. However, the associated error on this estimate increases dramatically, so it is unclear if this shift is statistically significant. We can confirm nevertheless that the value $\eta \sim1/3$ found in the agnostic approach is preferred over $\eta\sim1/2$ also in the restricted subsamples.

\bigskip

To summarise, the comparison between the results obtained by this \textit{biased} approach --~i.e. by imposing a cut in sphericity and centre offset~-- with the ones obtained with an \textit{agnostic} approach suggest that the scaling relations obtained by \citet{Schive14} may be not suitable to describe a cosmologically representative sample of dark matter haloes, since it includes a much more diverse population of systems for which the underlying assumptions are not valid.

\section{Conclusions}
\label{sec:conclusions}

With the ultimate goal of detailing the scaling relations that hold between the {\em solitonic} cores --~characterising the innermost regions of Fuzzy Dark Matter collapsed structures~-- and their hosting haloes properties in a realistic cosmological setup, we have developed a suite of cosmological simulations performed with the \AG code. The simulations span over a wide range of scales and environments, in order to understand the dependence of such relations on the formation history of individual haloes.\\

In the present work, which is the first of a series, we started from the most idealised situation of a single object forming from the collapse of a single quantum Jeans wavelength of the primordial FDM density field. With such simple and idealised test, we have confirmed the capability of the \AG code to reproduce the typical solitonic core of FDM systems that exhibits a density profile as described by Eq.~\ref{eq:soliton}: this result is clearly specific to \AG, however it confirms that the N-body approach is effective in the representation of the FDM framework.

We then moved to a less idealised and more cosmologically relevant scenario, in which we simulated eight individual haloes within their native cosmological environment by means of the zoom-in re-simulation technique. We found that every simulated halo forms a solitonic core and we confirm the general trend of haloes to feature a density profile that scales as $r^{-9/5}$ in the outer regions, far from the solitonic core, as observed also by \citet{Eggemeier19}.

Moreover, from the structural properties and the time evolution of our eight simulated haloes, we were able to estimate the main parameters (i.e. the normalisation factor and --~more importantly~-- the exponent) of the scaling relations that link the core density to the core radius and the core mass to the halo virial properties, which we termed SRI and SRII, respectively. We first performed a bootstrap sampling including all haloes at all redshifts independently from their structural, evolutionary, and environmental properties, finding that our sample is inconsistent with the SRI observed in \citet{Schive14}, while it is much more consistent with the SRII found in \citet{Mocz17} than the one observed in \citet{Schive14}.

Taking into account the dynamical state of the systems through the analysis of their sphericity and offset between centre of mass and the gravitational potential minimum, which we assumed to be a proxy for their relaxation, we demonstrated that unrelaxed systems (i.e. objects characterised by a low sphericity and a large centre offset) represent a subsample that systematically deviates from SRI and SRII of \citet{Schive14}, suggesting that the scaling relations presented in that work may be natively biased towards relaxed systems and not representative of a cosmological sample of haloes with a realistic distribution of relaxation states. In fact, by restricting the analysis only to a biased subsample of the most spherical and relaxed systems within our sample, it is possible to draw near to the expected values of the exponent of SRI found by \citet{Schive14} and of SRII by \citet{Mocz17}, thus supporting this claim. Given the fact that relaxed haloes are only a fraction of the total halo population found in the Universe, scaling relations meant to describe all haloes independently of their dynamical state are expected to be different from the relaxed case. In this sense, the scaling relations provided in this work are a better --~yet still partial~-- representation of the relations between core and halo properties in a realistic cosmological sample.

The results obtained through the analysis of merger histories confirmed that the zoom-in halo sample is biased towards weakly interacting systems, due to our selection procedure, therefore preventing a solid quantitative investigation of the impact of mergers on the onset of scaling relations. Nevertheless, mergers are the most effective way to alter the dynamical state of a systems and disrupt relaxation, thus they may play an important role in changing the effective scaling relations characterising FDM systems. In the next entry of this series \PARTII, we will address this topic with large cosmological simulations at fixed resolution, providing a close study of mergers in a cosmological context.

\bigskip

To summarise, with the present work we initiated a series of papers aiming at investigating the scaling relations arising between solitonic cores properties and their host haloes in FDM cosmologies. With the use of \AG, we built a sample of FDM haloes fully integrated in a common cosmological environment, simulated through the zoom-in approach. In particular, we focused on the impact of the morphological and dynamical state of the systems on the scaling relations previously found in the literature, which rely on the assumptions of sphericity and dynamical relaxation. We found that such scaling relations are not generally valid for our cosmological sample of haloes, as it includes systems that do not satisfy these conditions. These un-spherical and un-relaxed systems, which are legitimately part of a cosmological sample nevertheless, are responsible for the statistical deviation from the scaling relations obtained in the literature, thereby suggesting that these scaling relations are only valid for highly idealised systems and must be corrected for a realistic cosmological sample of haloes. We also find suggestions of an important role of merger events in the modifications of the scaling relations, that will be investigated in detail in an upcoming companion paper.

\section*{Acknowledgements}
The authors acknowledge support from the Italian Ministry for Education, University and Research (MIUR) through the SIR individual grant SIMCODE, project number {RBSI14P4IH}. The simulations presented in this work were performed on the Marconi supercomputer at CINECA, for which the authors acknowledge the ISCRA initiative for the availability of resources and support (two class C allocations, project numbers: HP10CPVK72 and HP10C0JGGH). This research was also supported by the Munich Institute for Astro- and Particle Physics (MIAPP) which is funded by the Deutsche Forschungsgemeinschaft (DFG, German Research Foundation) under Germany's Excellence Strategy - EXC-2094 - 390783311. The authors thank Philip Mocz and Andrea V. Macciò for useful comments and suggestions. MN is particularly grateful to Bodo Schwabe and Joshua Eby for the stimulating discussions during the Axion Cosmology programme at MIAPP and for further comments on the final form of this work. MN is also thankful to Cecilia Bacchini for her feedback in the early stages of this work and her incredible talent for acronyms.




\bibliographystyle{mnras}
\bibliography{BIB,baldi_bibliography} 

\begin{thebibliography}{}
\makeatletter
\relax
\def\mn@urlcharsother{\let\do\@makeother \do\$\do\&\do\#\do\^\do\_\do\%\do\~}
\def\mn@doi{\begingroup\mn@urlcharsother \@ifnextchar [ {\mn@doi@}
  {\mn@doi@[]}}
\def\mn@doi@[#1]#2{\def\@tempa{#1}\ifx\@tempa\@empty \href
  {http://dx.doi.org/#2} {doi:#2}\else \href {http://dx.doi.org/#2} {#1}\fi
  \endgroup}
\def\mn@eprint#1#2{\mn@eprint@#1:#2::\@nil}
\def\mn@eprint@arXiv#1{\href {http://arxiv.org/abs/#1} {{\tt arXiv:#1}}}
\def\mn@eprint@dblp#1{\href {http://dblp.uni-trier.de/rec/bibtex/#1.xml}
  {dblp:#1}}
\def\mn@eprint@#1:#2:#3:#4\@nil{\def\@tempa {#1}\def\@tempb {#2}\def\@tempc
  {#3}\ifx \@tempc \@empty \let \@tempc \@tempb \let \@tempb \@tempa \fi \ifx
  \@tempb \@empty \def\@tempb {arXiv}\fi \@ifundefined
  {mn@eprint@\@tempb}{\@tempb:\@tempc}{\expandafter \expandafter \csname
  mn@eprint@\@tempb\endcsname \expandafter{\@tempc}}}

\bibitem[\protect\citeauthoryear{Aghanim et~al.}{Aghanim
  et~al.}{2018b}]{Planck18}
Aghanim N.,  et~al., 2018b

\bibitem[\protect\citeauthoryear{Aghanim et~al.}{Aghanim
  et~al.}{2018a}]{Planck_2018_Gravitional_Lensing}
Aghanim N.,  et~al., 2018a

\bibitem[\protect\citeauthoryear{Albert et~al.}{Albert
  et~al.}{2017}]{Fermi17annih}
Albert A.,  et~al., 2017, \mn@doi [Astrophys. J.]
  {10.3847/1538-4357/834/2/110}, 834, 110

\bibitem[\protect\citeauthoryear{Armengaud, Palanque-Delabrouille, Marsh, Baur
  \& Y\`eche}{Armengaud et~al.}{2017}]{Armengaud17}
Armengaud E.,  Palanque-Delabrouille N.,  Marsh D. J.~E.,  Baur J.,   Y\`eche
  C.,  2017, \mn@doi [Mon. Not. Roy. Astron. Soc.] {10.1093/mnras/stx1870},
  471, 4606

\bibitem[\protect\citeauthoryear{Arvanitaki \& Geraci}{Arvanitaki \&
  Geraci}{2014}]{ARIADNE}
Arvanitaki A.,  Geraci A.~A.,  2014, \mn@doi [Phys. Rev. Lett.]
  {10.1103/PhysRevLett.113.161801}, 113, 161801

\bibitem[\protect\citeauthoryear{Banerjee, Eby, Kim, Matsedonskyi  \&
  Perez}{Banerjee et~al.}{2019}]{Banerjee19}
Banerjee A.,  Eby J.,  Kim H.,  Matsedonskyi O.,   Perez G.,  2019

\bibitem[\protect\citeauthoryear{Bar, Blas, Blum  \& Sibiryakov}{Bar
  et~al.}{2018}]{Bar18}
Bar N.,  Blas D.,  Blum K.,   Sibiryakov S.,  2018, \mn@doi [Phys. Rev. D]
  {10.1103/PhysRevD.98.083027}, 98, 083027

\bibitem[\protect\citeauthoryear{Bar, Blum, Eby  \& Sato}{Bar
  et~al.}{2019}]{Bar19}
Bar N.,  Blum K.,  Eby J.,   Sato R.,  2019, \mn@doi [Phys. Rev. D]
  {10.1103/PhysRevD.99.103020}, 99, 103020

\bibitem[\protect\citeauthoryear{Bertone, Hooper  \& Silk}{Bertone
  et~al.}{2005}]{Bertone_Hooper_Silk_2005}
Bertone G.,  Hooper D.,   Silk J.,  2005, \mn@doi [Phys. Rept.]
  {10.1016/j.physrep.2004.08.031}, 405, 279

\bibitem[\protect\citeauthoryear{Bode, Ostriker  \& Turok}{Bode
  et~al.}{2001}]{Bode00}
Bode P.,  Ostriker J.~P.,   Turok N.,  2001, \mn@doi [Astrophys. J.]
  {10.1086/321541}, 556, 93

\bibitem[\protect\citeauthoryear{{Bohm}}{{Bohm}}{1952}]{Bohm52}
{Bohm} D.,  1952, \mn@doi [Physical Review] {10.1103/PhysRev.85.166}, \href
  {http://adsabs.harvard.edu/abs/1952PhRv...85..166B} {85, 166}

\bibitem[\protect\citeauthoryear{Bosma}{Bosma}{1981}]{Bosma_1981}
Bosma A.,  1981, Astron. J., 86, 1825

\bibitem[\protect\citeauthoryear{Braine et~al.,}{Braine et~al.}{2020}]{ADMX20}
Braine T.,  et~al., 2020, \mn@doi [Phys. Rev. Lett.]
  {10.1103/PhysRevLett.124.101303}, 124, 101303

\bibitem[\protect\citeauthoryear{Bryan \& Norman}{Bryan \&
  Norman}{1998}]{Bryan97}
Bryan G.~L.,  Norman M.~L.,  1998, \mn@doi [Astrophys. J.] {10.1086/305262},
  495, 80

\bibitem[\protect\citeauthoryear{Buonaura}{Buonaura}{2018}]{Buonaura18}
Buonaura A.,  2018, \mn@doi [PoS] {10.22323/1.297.0079}, DIS2017, 079

\bibitem[\protect\citeauthoryear{{Chavanis}}{{Chavanis}}{2011}]{Chavanis11a}
{Chavanis} P.-H.,  2011, \mn@doi [\prd] {10.1103/PhysRevD.84.043531}, \href
  {https://ui.adsabs.harvard.edu/abs/2011PhRvD..84d3531C} {84, 043531}

\bibitem[\protect\citeauthoryear{{Chavanis} \& {Delfini}}{{Chavanis} \&
  {Delfini}}{2011}]{Chavanis11b}
{Chavanis} P.-H.,  {Delfini} L.,  2011, \mn@doi [\prd]
  {10.1103/PhysRevD.84.043532}, \href
  {https://ui.adsabs.harvard.edu/abs/2011PhRvD..84d3532C} {84, 043532}

\bibitem[\protect\citeauthoryear{Clowe, Bradac, Gonzalez, Markevitch, Randall,
  Jones  \& Zaritsky}{Clowe et~al.}{2006}]{Clowe06}
Clowe D.,  Bradac M.,  Gonzalez A.~H.,  Markevitch M.,  Randall S.~W.,  Jones
  C.,   Zaritsky D.,  2006, \mn@doi [Astrophys. J.] {10.1086/508162}, 648, L109

\bibitem[\protect\citeauthoryear{Danninger}{Danninger}{2017}]{Danninger17}
Danninger M.,  2017, \mn@doi [J. Phys. Conf. Ser.]
  {10.1088/1742-6596/888/1/012039}, 888, 012039

\bibitem[\protect\citeauthoryear{{Davis}, {Efstathiou}, {Frenk}  \&
  White}{{Davis} et~al.}{1985}]{Davis_etal_1985}
{Davis} M.,  {Efstathiou} G.,  {Frenk} C.~S.,   White S.~D.,  1985, \mn@doi
  [Astrophys.J.] {10.1086/163168}, 292, 371

\bibitem[\protect\citeauthoryear{Desjacques \& Nusser}{Desjacques \&
  Nusser}{2019}]{Desjacques19}
Desjacques V.,  Nusser A.,  2019, \mn@doi [Mon. Not. Roy. Astron. Soc.]
  {10.1093/mnras/stz1978}, 488, 4497

\bibitem[\protect\citeauthoryear{Du, Behrens, Niemeyer  \& Schwabe}{Du
  et~al.}{2017}]{Du16}
Du X.,  Behrens C.,  Niemeyer J.~C.,   Schwabe B.,  2017, \mn@doi [Phys. Rev.]
  {10.1103/PhysRevD.95.043519}, D95, 043519

\bibitem[\protect\citeauthoryear{Eggemeier \& Niemeyer}{Eggemeier \&
  Niemeyer}{2019}]{Eggemeier19}
Eggemeier B.,  Niemeyer J.~C.,  2019, \mn@doi [Phys. Rev.]
  {10.1103/PhysRevD.100.063528}, D100, 063528

\bibitem[\protect\citeauthoryear{Feng}{Feng}{2010}]{Feng10}
Feng J.~L.,  2010, \mn@doi [Annual Review of Astronomy and Astrophysics]
  {10.1146/annurev-astro-082708-101659}, 48, 495

\bibitem[\protect\citeauthoryear{Ferreira}{Ferreira}{2020}]{Ferreira20review}
Ferreira E.~G.,  2020

\bibitem[\protect\citeauthoryear{Graham \& Rajendran}{Graham \&
  Rajendran}{2013}]{CASPER}
Graham P.~W.,  Rajendran S.,  2013, \mn@doi [Phys. Rev. D]
  {10.1103/PhysRevD.88.035023}, 88, 035023

\bibitem[\protect\citeauthoryear{Gross}{Gross}{1961}]{Gross61}
Gross E.~P.,  1961, \mn@doi [Il Nuovo Cimento] {10.1007/BF02731494}, \href
  {http://adsabs.harvard.edu/abs/1961NCim...20..454G} {20, 454}

\bibitem[\protect\citeauthoryear{{Hahn} \& {Abel}}{{Hahn} \&
  {Abel}}{2011}]{music}
{Hahn} O.,  {Abel} T.,  2011, \mn@doi [\mnras]
  {10.1111/j.1365-2966.2011.18820.x}, \href
  {https://ui.adsabs.harvard.edu/abs/2011MNRAS.415.2101H} {415, 2101}

\bibitem[\protect\citeauthoryear{{Heymans} et~al.,}{{Heymans}
  et~al.}{2013}]{Heymans_etal_2013}
{Heymans} C.,  et~al., 2013, \mn@doi [\mnras] {10.1093/mnras/stt601}, \href
  {http://adsabs.harvard.edu/abs/2013MNRAS.432.2433H} {432, 2433}

\bibitem[\protect\citeauthoryear{{Hildebrandt} et~al.,}{{Hildebrandt}
  et~al.}{2017}]{Hildebrandt_etal_2017}
{Hildebrandt} H.,  et~al., 2017, \mn@doi [\mnras] {10.1093/mnras/stw2805},
  \href {http://adsabs.harvard.edu/abs/2017MNRAS.465.1454H} {465, 1454}

\bibitem[\protect\citeauthoryear{Hlozek, Grin, Marsh  \& Ferreira}{Hlozek
  et~al.}{2015}]{axionCAMB}
Hlozek R.,  Grin D.,  Marsh D. J.~E.,   Ferreira P.~G.,  2015, \mn@doi [Phys.
  Rev.] {10.1103/PhysRevD.91.103512}, D91, 103512

\bibitem[\protect\citeauthoryear{Hu, Barkana  \& Gruzinov}{Hu
  et~al.}{2000}]{Hu00}
Hu W.,  Barkana R.,   Gruzinov A.,  2000, \mn@doi [Phys. Rev. Lett.]
  {10.1103/PhysRevLett.85.1158}, 85, 1158

\bibitem[\protect\citeauthoryear{Hui, Ostriker, Tremaine  \& Witten}{Hui
  et~al.}{2017}]{Hui16}
Hui L.,  Ostriker J.~P.,  Tremaine S.,   Witten E.,  2017, \mn@doi [Phys. Rev.]
  {10.1103/PhysRevD.95.043541}, D95, 043541

\bibitem[\protect\citeauthoryear{Ir\v{s}i\v{c}, Viel, Haehnelt, Bolton  \&
  Becker}{Ir\v{s}i\v{c} et~al.}{2017}]{Irsic17}
Ir\v{s}i\v{c} V.,  Viel M.,  Haehnelt M.~G.,  Bolton J.~S.,   Becker G.~D.,
  2017, \mn@doi [Phys. Rev. Lett.] {10.1103/PhysRevLett.119.031302}, 119,
  031302

\bibitem[\protect\citeauthoryear{Ji \& Sin}{Ji \& Sin}{1994}]{Ji94}
Ji S.~U.,  Sin S.~J.,  1994, \mn@doi [Phys. Rev.] {10.1103/PhysRevD.50.3655},
  D50, 3655

\bibitem[\protect\citeauthoryear{Jungman, Kamionkowski  \& Griest}{Jungman
  et~al.}{1996}]{Jungman95}
Jungman G.,  Kamionkowski M.,   Griest K.,  1996, \mn@doi [Phys. Rept.]
  {10.1016/0370-1573(95)00058-5}, 267, 195

\bibitem[\protect\citeauthoryear{Katz, Quinn, Bertschinger  \& Gelb}{Katz
  et~al.}{1994}]{Katz94}
Katz N.,  Quinn T.,  Bertschinger E.,   Gelb J.~M.,  1994, \mn@doi [Monthly
  Notices of the Royal Astronomical Society] {10.1093/mnras/270.1.L71}, 270,
  L71

\bibitem[\protect\citeauthoryear{Klypin, Kravtsov, Valenzuela  \& Prada}{Klypin
  et~al.}{1999}]{Klypin_etal_1999}
Klypin A.~A.,  Kravtsov A.~V.,  Valenzuela O.,   Prada F.,  1999, \mn@doi
  [Astrophys.J.] {10.1086/307643}, 522, 82

\bibitem[\protect\citeauthoryear{Kolb \& Tkachev}{Kolb \&
  Tkachev}{1993}]{Kolb93}
Kolb E.~W.,  Tkachev I.~I.,  1993, \mn@doi [Phys. Rev. Lett.]
  {10.1103/PhysRevLett.71.3051}, 71, 3051

\bibitem[\protect\citeauthoryear{Kolb \& Tkachev}{Kolb \&
  Tkachev}{1994}]{Kolb94}
Kolb E.~W.,  Tkachev I.~I.,  1994, \mn@doi [Phys. Rev. D]
  {10.1103/PhysRevD.49.5040}, 49, 5040

\bibitem[\protect\citeauthoryear{Komatsu et~al.}{Komatsu et~al.}{2011}]{wmap7}
Komatsu E.,  et~al., 2011, \mn@doi [Astrophys. J. Suppl.]
  {10.1088/0067-0049/192/2/18}, 192, 18

\bibitem[\protect\citeauthoryear{Kuhlen, Vogelsberger  \& Angulo}{Kuhlen
  et~al.}{2012}]{Kuhlen_Vogelsberger_Angulo_2012}
Kuhlen M.,  Vogelsberger M.,   Angulo R.,  2012, arXiv:1209.5745

\bibitem[\protect\citeauthoryear{Laguë, Bond, Hlo\v~zek, Marsh  \&
  Söding}{Laguë et~al.}{2020}]{Lague20}
Laguë A.,  Bond J.~R.,  Hlo\v~zek R.,  Marsh D.~J.,   Söding L.,  2020

\bibitem[\protect\citeauthoryear{Levkov, Panin  \& Tkachev}{Levkov
  et~al.}{2018}]{Levkov18}
Levkov D.~G.,  Panin A.~G.,   Tkachev I.~I.,  2018, \mn@doi [Phys. Rev. Lett.]
  {10.1103/PhysRevLett.121.151301}, 121, 151301

\bibitem[\protect\citeauthoryear{{Macci{\`o}}, {Dutton}  \& {van den
  Bosch}}{{Macci{\`o}} et~al.}{2008}]{Maccio_etal_2008}
{Macci{\`o}} A.~V.,  {Dutton} A.~A.,   {van den Bosch} F.~C.,  2008, \mn@doi
  [\mnras] {10.1111/j.1365-2966.2008.14029.x}, \href
  {http://adsabs.harvard.edu/abs/2008MNRAS.391.1940M} {391, 1940}

\bibitem[\protect\citeauthoryear{Madelung}{Madelung}{1927}]{Madelung27}
Madelung E.,  1927, \mn@doi [Zeitschrift f{\"u}r Physik] {10.1007/BF01400372},
  40, 322

\bibitem[\protect\citeauthoryear{Majorovits et~al.}{Majorovits
  et~al.}{2020}]{MADMAX17}
Majorovits B.,  et~al., 2020, \mn@doi [J. Phys. Conf. Ser.]
  {10.1088/1742-6596/1342/1/012098}, 1342, 012098

\bibitem[\protect\citeauthoryear{Marsh}{Marsh}{2016}]{Marsh16review}
Marsh D. J.~E.,  2016, \mn@doi [Phys. Rept.] {10.1016/j.physrep.2016.06.005},
  643, 1

\bibitem[\protect\citeauthoryear{Mateo}{Mateo}{1998}]{Mateo98}
Mateo M.,  1998, \mn@doi [Ann. Rev. Astron. Astrophys.]
  {10.1146/annurev.astro.36.1.435}, 36, 435

\bibitem[\protect\citeauthoryear{McAllister, Flower, Ivanov, Goryachev,
  Bourhill  \& Tobar}{McAllister et~al.}{2017}]{ORGAN}
McAllister B.~T.,  Flower G.,  Ivanov E.~N.,  Goryachev M.,  Bourhill J.,
  Tobar M.~E.,  2017, \mn@doi [Phys. Dark Univ.] {10.1016/j.dark.2017.09.010},
  18, 67

\bibitem[\protect\citeauthoryear{Mocz \& Succi}{Mocz \& Succi}{2015}]{Mocz15}
Mocz P.,  Succi S.,  2015, \mn@doi [Phys. Rev.] {10.1103/PhysRevE.91.053304},
  E91, 053304

\bibitem[\protect\citeauthoryear{Mocz, Vogelsberger, Robles, Zavala,
  Boylan-Kolchin, Fialkov  \& Hernquist}{Mocz et~al.}{2017}]{Mocz17}
Mocz P.,  Vogelsberger M.,  Robles V.~H.,  Zavala J.,  Boylan-Kolchin M.,
  Fialkov A.,   Hernquist L.,  2017, \mn@doi [Mon. Not. Roy. Astron. Soc.]
  {10.1093/mnras/stx1887}, 471, 4559

\bibitem[\protect\citeauthoryear{{Mocz} et~al.,}{{Mocz} et~al.}{2019}]{Mocz19}
{Mocz} P.,  et~al., 2019, \mn@doi [\prl] {10.1103/PhysRevLett.123.141301},
  \href {https://ui.adsabs.harvard.edu/abs/2019PhRvL.123n1301M} {123, 141301}

\bibitem[\protect\citeauthoryear{{Mocz} et~al.,}{{Mocz}
  et~al.}{2020}]{Mocz19companion}
{Mocz} P.,  et~al., 2020, \mn@doi [\mnras] {10.1093/mnras/staa738}, \href
  {https://ui.adsabs.harvard.edu/abs/2020MNRAS.494.2027M} {494, 2027}

\bibitem[\protect\citeauthoryear{Navarro \& White}{Navarro \&
  White}{1994}]{Navarro94}
Navarro J.~F.,  White S. D.~M.,  1994, \mn@doi [Monthly Notices of the Royal
  Astronomical Society] {10.1093/mnras/267.2.401}, 267, 401

\bibitem[\protect\citeauthoryear{Neto et~al.}{Neto
  et~al.}{2007}]{Neto_etal_2007}
Neto A.~F.,  et~al., 2007

\bibitem[\protect\citeauthoryear{Nori \& Baldi}{Nori \& Baldi}{2018}]{Nori18}
Nori M.,  Baldi M.,  2018, \mn@doi [Mon. Not. Roy. Astron. Soc.]
  {10.1093/mnras/sty1224}, 478, 3935

\bibitem[\protect\citeauthoryear{Nori, Murgia, Ir\v{s}i\v{c}, Baldi  \&
  Viel}{Nori et~al.}{2019}]{Nori19}
Nori M.,  Murgia R.,  Ir\v{s}i\v{c} V.,  Baldi M.,   Viel M.,  2019, \mn@doi
  [Mon. Not. Roy. Astron. Soc.] {10.1093/mnras/sty2888}, 482, 3227

\bibitem[\protect\citeauthoryear{Oh, de Blok, Brinks, Walter  \& Kennicutt}{Oh
  et~al.}{2011}]{Oh11}
Oh S.-H.,  de Blok W. J.~G.,  Brinks E.,  Walter F.,   Kennicutt Jr R.~C.,
  2011, \mn@doi [Astron. J.] {10.1088/0004-6256/141/6/193}, 141, 193

\bibitem[\protect\citeauthoryear{Ouellet et~al.}{Ouellet
  et~al.}{2019}]{ABRACADABRA19}
Ouellet J.~L.,  et~al., 2019, \mn@doi [Phys. Rev. D]
  {10.1103/PhysRevD.99.052012}, 99, 052012

\bibitem[\protect\citeauthoryear{Peccei \& Quinn}{Peccei \&
  Quinn}{1977a}]{PecceiQuinn77b}
Peccei R.~D.,  Quinn H.~R.,  1977a, \mn@doi [Phys. Rev. D]
  {10.1103/PhysRevD.16.1791}, 16, 1791

\bibitem[\protect\citeauthoryear{Peccei \& Quinn}{Peccei \&
  Quinn}{1977b}]{PecceiQuinn77a}
Peccei R.~D.,  Quinn H.~R.,  1977b, \mn@doi [Phys. Rev. Lett.]
  {10.1103/PhysRevLett.38.1440}, 38, 1440

\bibitem[\protect\citeauthoryear{Peebles}{Peebles}{1980}]{Peebles80}
Peebles P. J.~E.,  1980, The large-scale structure of the universe.
Princeton University Press Princeton, N.J

\bibitem[\protect\citeauthoryear{Persic, Salucci  \& Stel}{Persic
  et~al.}{1996}]{Persic96}
Persic M.,  Salucci P.,   Stel F.,  1996, \mn@doi [Mon. Not. Roy. Astron. Soc.]
  {10.1093/mnras/281.1.27, 10.1093/mnras/278.1.27}, 281, 27

\bibitem[\protect\citeauthoryear{Pitaevskii}{Pitaevskii}{1961}]{Pitaevskii61}
Pitaevskii L.~P.,  1961, \mn@doi [Sov. Phys. JETP] {10.1007/BF02731494}, \href
  {http://adsabs.harvard.edu/abs/1961NCim...20..454G} {13, 451}

\bibitem[\protect\citeauthoryear{{Rubin}, {Ford}  \& {.~Thonnard}}{{Rubin}
  et~al.}{1980}]{Rubin_Ford_Thonnard_1980}
{Rubin} V.~C.,  {Ford} W.~K.~J.,   {.~Thonnard} N.,  1980, \mn@doi [\apj]
  {10.1086/158003}, \href {http://adsabs.harvard.edu/abs/1980ApJ...238..471R}
  {238, 471}

\bibitem[\protect\citeauthoryear{Schive, Tsai  \& Chiueh}{Schive
  et~al.}{2010}]{GAMER}
Schive H.-Y.,  Tsai Y.-C.,   Chiueh T.,  2010, \mn@doi [Astrophys. J. Suppl.]
  {10.1088/0067-0049/186/2/457}, 186, 457

\bibitem[\protect\citeauthoryear{Schive, Chiueh  \& Broadhurst}{Schive
  et~al.}{2014}]{Schive14}
Schive H.-Y.,  Chiueh T.,   Broadhurst T.,  2014, \mn@doi [Nature Phys.]
  {10.1038/nphys2996}, 10, 496

\bibitem[\protect\citeauthoryear{Schive, Chiueh, Broadhurst  \& Huang}{Schive
  et~al.}{2016}]{Schive16}
Schive H.-Y.,  Chiueh T.,  Broadhurst T.,   Huang K.-W.,  2016, \mn@doi
  [Astrophys. J.] {10.3847/0004-637X/818/1/89}, 818, 89

\bibitem[\protect\citeauthoryear{Schive, ZuHone, Goldbaum, Turk, Gaspari  \&
  Cheng}{Schive et~al.}{2017}]{GAMER2}
Schive H.-Y.,  ZuHone J.~A.,  Goldbaum N.~J.,  Turk M.~J.,  Gaspari M.,   Cheng
  C.-Y.,  2017

\bibitem[\protect\citeauthoryear{Schwabe, Niemeyer  \& Engels}{Schwabe
  et~al.}{2016}]{Schwabe16}
Schwabe B.,  Niemeyer J.~C.,   Engels J.~F.,  2016, \mn@doi [Phys. Rev.]
  {10.1103/PhysRevD.94.043513}, D94, 043513

\bibitem[\protect\citeauthoryear{Sikivie}{Sikivie}{2008}]{Sikivie08}
Sikivie P.,  2008, \mn@doi [Lect. Notes Phys.] {10.1007/978-3-540-73518-2\_2},
  741, 19

\bibitem[\protect\citeauthoryear{Springel}{Springel}{2005}]{Springel05}
Springel V.,  2005, \mn@doi [Mon. Not. Roy. Astron. Soc.]
  {10.1111/j.1365-2966.2005.09655.x}, 364, 1105

\bibitem[\protect\citeauthoryear{{Springel}, {White}, {Tormen}  \&
  {Kauffmann}}{{Springel} et~al.}{2001}]{Springel_etal_2001}
{Springel} V.,  {White} S.~D.~M.,  {Tormen} G.,   {Kauffmann} G.,  2001,
  \mn@doi [\mnras] {10.1046/j.1365-8711.2001.04912.x}, \href
  {http://adsabs.harvard.edu/abs/2001MNRAS.328..726S} {328, 726}

\bibitem[\protect\citeauthoryear{{Springel} et~al.,}{{Springel}
  et~al.}{2008}]{Aquarius}
{Springel} V.,  et~al., 2008, \mn@doi [\mnras]
  {10.1111/j.1365-2966.2008.14066.x}, \href
  {http://adsabs.harvard.edu/abs/2008MNRAS.391.1685S} {391, 1685}

\bibitem[\protect\citeauthoryear{Turk, Smith, Oishi, Skory, Skillman, Abel  \&
  Norman}{Turk et~al.}{2011}]{YT}
Turk M.~J.,  Smith B.~D.,  Oishi J.~S.,  Skory S.,  Skillman S.~W.,  Abel T.,
  Norman M.~L.,  2011, \mn@doi [The Astrophysical Journal Supplement Series]
  {10.1088/0067-0049/192/1/9}, 192, 9

\bibitem[\protect\citeauthoryear{Veltmaat, Niemeyer  \& Schwabe}{Veltmaat
  et~al.}{2018}]{Veltmaat18}
Veltmaat J.,  Niemeyer J.~C.,   Schwabe B.,  2018, \mn@doi [Phys. Rev.]
  {10.1103/PhysRevD.98.043509}, D98, 043509

\bibitem[\protect\citeauthoryear{Woo \& Chiueh}{Woo \& Chiueh}{2009}]{Woo09}
Woo T.-P.,  Chiueh T.,  2009, The Astrophysical Journal, 697, 850

\bibitem[\protect\citeauthoryear{Zel'dovich}{Zel'dovich}{1970}]{Zeldovich_1970}
Zel'dovich Y.~B.,  1970, Astron. Astrophys., 5, 84

\bibitem[\protect\citeauthoryear{Zwicky}{Zwicky}{1937}]{Zwicky_1937}
Zwicky F.,  1937, Astrophys. J., 86, 217

\makeatother
\end{thebibliography}



\appendix

\section{Absorbing the scale factor and boson mass dependences}
\label{app:scalefactor}

The EP system described in Eq.~\ref{eq:EP} can be recast in a scale factor $a$ and boson mass $m_{\chi}$ independent fashion through the coordinate transformation
\begin{alignat}{2}
\label{eq:ascaling}
& \begin{dcases}
x = \tilde x \ a^{\epsilon} \ m_{\chi}^{\delta} \\
t = \tilde t \ a^{2+2\epsilon} \ m_{\chi}^{1+2\delta}\\
u = \tilde u \ a^{-2 -\epsilon} \ m_{\chi}^{-1-\delta}\\
\rho = \tilde \rho \ a^{-1 -4\epsilon} \ m_{\chi}^{-2-4\delta}\\
\end{dcases}
& \qquad & \begin{dcases}
M = \tilde M \ a^{-1 -\epsilon} \ m_{\chi}^{-2-\delta} \\
\Phi = \tilde \Phi \ a^{-2 - 2\epsilon} \ m_{\chi}^{-2-2\delta} \\
E = \tilde E \ a^{-5 -3\epsilon} \ m_{\chi}^{-4-3\delta}
\end{dcases}
\end{alignat}

where $\epsilon$ and $\delta$ are parameters of choice. Note that this is the more general form of Eq.~\ref{eq:transformation}, where any dependence of $\lambda$ on scale factor or boson mass $\lambda \propto a^\epsilon m_{\chi}^\delta$ is here taken into account. It is easy to see that the choice of $\epsilon$ (as it was for the parameter $\lambda$) is completely irrelevant to the validity of any scaling relation presented in Sec.~\ref{sec:fdm_sr}, as long as Eq.~\ref{eq:EP} is valid. In fact, both SRI and SRII are $a$ and $m{\chi}$ independent when expressed in the tilded coordinates. In this work, we will adopt $\epsilon=\delta=0$, meaning that we do not apply any transformation to the comoving frame; as a comparison,  $\epsilon=-1/4$ was used in \citet{Schive14}, so that $\tilde \rho / \rho$ is not a function of time. It is interesting to note also that $\epsilon=-1$ maps the comoving frame into the physical one.

The assumption of constant scale factor in time $a \neq a(t)$ is obviously not valid in a cosmological scenario. In this case, the terms involving $\dot a = a H$ originating by the the transformation cannot be neglected, effectively breaking the scale factor independence. It is possible to neglect such derivatives assuming that the time for the solitonic solution to form $t_{\rm sol}$ is much shorter than the cosmic expansion time-scale $t_{\rm sol} \ll 1/H$, so that a solution to Eq.~\ref{eq:EP} can form on a much faster time-scale in which the scale factor can be regarded as almost constant \citep[see][for analyses on time evolution of solitons]{Levkov18}.

\section{From cylindrical to spherical soliton}
\label{sec:filament}

\begin{figure}
\includegraphics[width=\columnwidth, trim={0.5cm 0.5cm 0.5cm 0.5cm} ,clip]{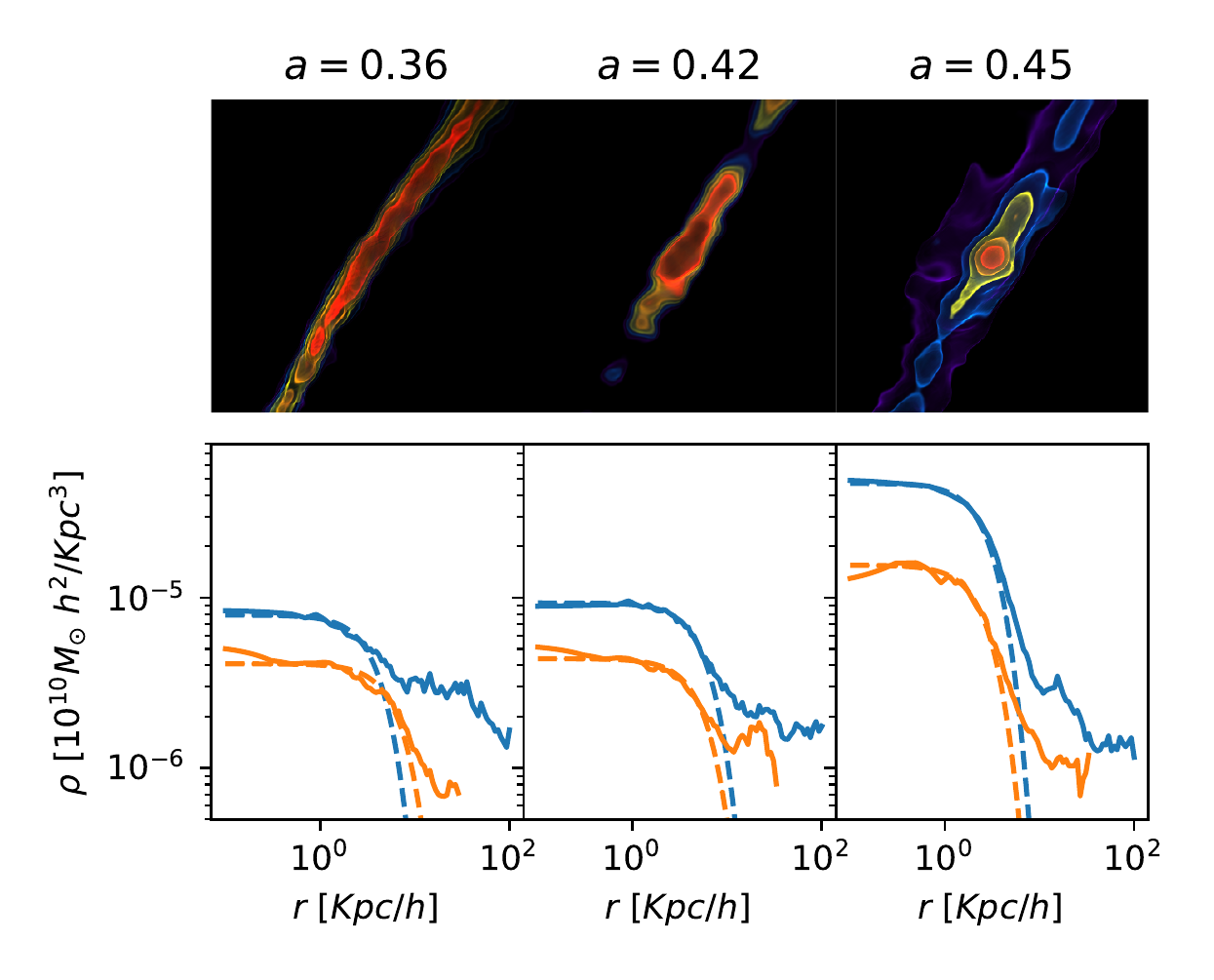}
\caption{3D rendering (top panels) and radial density profiles (bottom panels) of the H halo at different times. The rendering represent a $50 \ \dimR$ cube where colors for density levels are chosen consistently with Fig.~\ref{fig:zoom_cores}. Spherical (blue) and cylindrical (orange) radial profiles (solid lines) are plotted together with their corresponding core profiles (dashed lines). Radial profiles share the same plot for visual purposes: note, however, that the x-axis represent the \textit{spherical} and the \textit{cylindrical} radial coordinate, for the former and the latter case respectively.}
\label{fig:filament}
\end{figure}

A peculiar case worth mentioning is represented by the \textit{H} zoom-in halo: the smallest halo of our set forms via cylindrical collapse of a filament around $z=1.8$ and sets into a spherical configuration around $z=1.2$. The collapse of a filament and the consequential formation of a "core" --~though the term "core" in this case may sound dissonant with the cylindrical shape of the system~-- was also recently observed in a simulation in \citet{Mocz19,Mocz19companion}. The authors provided a numerical approximation for the cylindrical solution equivalent to Eq.~\ref{eq:soliton}, finding the same functional form but with a constant $\alpha=0.127$.

In Fig.~\ref{fig:filament}, we present the 3D rendering of the density distribution (upper panels), in a cube of side $50\ \dimR$ during the transition from cylindrical to spherical symmetry, with the corresponding radial density profiles (lower panels). The color scheme of the 3D rendering are the same of Fig.~\ref{fig:zoom_cores} for consistency. For each redshift, the radial profile is shown as computed on spherical shells and fitting the core using Eq.~\ref{eq:soliton} (solid and dashed blue lines) as well as obtained considering cylindrical shells and using the cylindrical version of the core equation suggested by \citet{Mocz19} (solid and dashed orange lines). Technically, the cylindrical density profile is computed on radial cylindrical shells generated around the major semi-axis $a$ which is taken as the axis of symmetry; as the profile is computed on particles belonging to the halo, the longitudinal extent is limited by the farthest particle from the halo centre. Note that, although presented in the same plot for visual purposes, the distance in the x-axis take different meaning for the two observables, representing the spherical radius and the cylindrical radius, respectively.

It is qualitatively interesting that the cylindrical profile exhibits the presence of a "core" from the beginning, while this feature emerges only at the end of the transition in the spherical profile. For the first time in the literature, --~to the best of our knowledge~-- we presented here the transition between a cylindrical and spherical regime of a FDM core; more generally, this system represents an example of the complex FDM halo evolution that can take place in filaments, that can have interesting astrophysical implications \citep[as suggested by][]{Mocz19,Mocz19companion}.


\label{lastpage}
\end{document}